\documentclass[preprint,12pt]{elsarticle}

\usepackage{graphics}
\usepackage{subfigure}
\usepackage{amssymb}

\journal{Journal of Computational Physics}

\begin{document}

\begin{frontmatter}

\title{A simulation method for determining the optical response of highly complex photonic structures of biological origin}

\author[label1,label2]{A.E. Dolinko}
\ead{adolinko@df.uba.ar}
\author[label1]{D.C. Skigin}

\address[label1]{Grupo de Electromagnetismo Aplicado, Departamento de F´isica, FCEN, Universidad de Buenos Aires, and IFIBA - CONICET, Ciudad Universitaria, Pabell´on I, C1428EHA Buenos Aires, Argentina.}
\address[label2]{Departamento de Biodiversidad y Biolog\'ia Experimental, Facultad de Ciencias Exactas y Naturales, Universidad de Buenos Aires, Ciudad Universitaria, Pabell\'on II, C1428EHA Buenos Aires, Argentina.}

\begin{abstract}
We present a method based on a time domain simulation
of wave propagation that allows studying the optical response of
a broad range of dielectric photonic structures. This method is
particularly suitable for dealing with complex biological
structures. One of the main features of the proposed approach is
the simple and intuitive way of defining the setup and the
photonic structure to be simulated, which can be done by feeding
the simulation with a digital image of the structure. We also
develop a set of techniques to process the behavior of the
evolving waves within the simulation. These techniques include a
\emph{direction filter}, that permits decoupling of waves
travelling simultaneously in different directions, a \emph{dynamic
differential absorber}, to cancel the waves reflected at the edges
of the simulation space, a multi-frequency excitation scheme based
on a filter that allows decoupling waves of different wavelengths
travelling simultaneously, and a near-to-far-field approach to
evaluate the resulting wavefield outside the simulation domain.
We validate the code and, as an example, apply it to the complex
structure found in a microorganism called {\em Diachea leucopoda},
which exhibits a multicolor iridescent appearance.
\end{abstract}


\end{frontmatter}

\section{Introduction}
\label{sec01}

The study of the optical response of structures with typical sizes of
the order of the optical wavelengths has gain great
interest in recent years. Emerging technologies had resulted
from the study of photonic materials, which consist of a regular
distribution of particles within a host
matrix. Depending on the size of the inclusions relative to the operating
wavelength, photonic materials can be classified in metamaterials
(characteristic size significantly smaller than the wavelength) and
photonic crystals (characteristic size of the order of the wavelength).
Then, metamaterials operating within the visible range are nano-structured dielectric or
metallic materials, which are designed to control the effective electric
permittivity and magnetic permeability in order to obtain a
specific response. In particular, the possibility of generating a
material with negative refraction index \cite{one} led to
a great variety of promising applications, such as
superlensing and optical cloaking \cite{two,three,four,five,six}.
On the other hand, all-dielectric photonic crystals exhibit interesting
properties that arise from the resonant scattering generated by the
specific modulation of the refraction index in a dielectric
transparent material \cite{seven,eight,nine}. Photons in this kind of medium play
the same role as electrons in the atomic crystal lattice
\cite{Joannopoulos,ten}, and then the photonic system also exhibits band gaps, i.e.,
forbidden bands for the propagation of waves. These artificial materials
can be designed for specific purposes such as optical switches, Bragg filters
or photonic crystal fibers \cite{eleven,twelve,thirteen}.

Another growing research field based on dielectric photonic
structures is the study of natural structural color, which is
responsible for the iridescent appearance that exhibit a broad diversity of
animals and plants \cite{fourteen,fifteen,sixteen}. Structural color
is produced by the selective reflection of light incident on the microscopic structures
present in the cover tissues of biological organisms.
Optical mechanisms such as interference, diffraction and scattering are
involved to achieve colorful patterns or metallic appearance. These
colors usually appear considerably brighter than those of
pigments, although they often result from completely transparent
materials \cite{seventeen,eighteen}. Unlike artificial
photonic materials, the geometry and distribution of these natural
media is usually extremely complex, and the simulation of their
electromagnetic response require versatile and accurate tools.
The study of this
phenomenon contributes to understand different behavioural functions
of living species such as thermoregulation and camouflage and, at the same time,
inspires new developments of artificial devices.

A large variety of rigorous electromagnetic methods to obtain the
optical response of a given photonic structure are available in
the literature. Among these methods, we can mention the modal
method \cite{Andrewartha,nineteen,twenty,twentyone,twenty-two},
coordinate-transformation methods \cite{twenty-three,twenty-four}
and the integral method
\cite{twenty-five,LesterSkigin,twenty-six}. These approaches are
very efficient for the accurate determination of the optical
response of corrugated interfaces and periodic gratings of
canonical shapes. However, in most cases this kind of methods are
not suitable for dealing with highly complex structures.

Another way of studying the electromagnetic response of complex
nano-structures is by means of computer simulations. A very spread
approach is the \emph{Finite Difference Time Domain}
method (FDTD) introduced by Taflove \emph{ et al.} around 40 years ago
\cite{twenty-seven}. This method is based on the Yee algorithm
\cite{twenty-eight} and consists in numerically solving
six coupled vector equations obtained from Maxwell's
equations in the time domain. The FDTD is a very powerful
method and has been improved during the last decades to account for a
great variety of problems in electrodynamics. However,
it results heavily time-consuming, and requires large
computer resources and even parallelization for very large simulation spaces
\cite{twenty-nine}.

In this paper, we present a very simple simulation method that
allows studying the propagation of electromagnetic waves in a
dielectric medium of arbitrary refraction index distribution. Due
to its simplicity, the evolution of the propagating waves can be
easily visualized on a conventional computer during
runtime. One of the highlights of the
proposed method is its versatility to obtain the optical
response of an arbitrary dielectric photonic structure, which can
be introduced within the simulation by means of a diagram or a
photograph in a form of a digital image or bitmap.
Then, the structure to be
studied can be easily generated by means of any photo-editor
software. In the case of natural photonic structures, an electron
microscope image of the real specimen can be used.

In Sec. 2 we summarize the basic concepts of the simulation.
Within the framework of the developed method, in Sections 3-6 we
present a set of techniques which permit us to control and analyze
the behavior of the waves within the simulation. We implemented a
\emph{direction filter} that permits decoupling waves travelling
simultaneously in different directions, and also allows
determining the field of energy flux in any type of wave (Sec. 3).
In Sec. 4 we present an active system to cancel waves
reflected at the edges of the simulation space, which allows
simulating boundary conditions that represent an unbounded virtual
space. This technique is independent of the wavelength and
also works for any waveform, and then, it is capable of handling
a multi-frequency excitation. Unlike other techniques to simulate
perfectly absorbing boundary conditions, the proposed approach
does not need to establish a finite thickness layer, and this
constitutes one of its main advantages since it avoids wasting
simulation space and time for that purpose. A multi-frequency
excitation scheme based on a filter that allows decoupling of
waves of several wavelengths travelling simultaneously is also
shown in Sec. 5. Such a system increases the computing speed since it avoids
making a sequential frequency sweep to obtain the spectral
response of a given structure. Finally, in Sec. 6 we show how to obtain the
far field (outside the simulation space) from the near field.
As an example, in Sec. 7 we apply the simulation method,
including the whole set of developed techniques,
to obtain the optical response of the photonic structure
present in the tissue of a microorganism. Concluding remarks are
given in Section 8.

\section{Description of the simulation}
\label{sec02}

\begin{figure}
\centering
\includegraphics[width=9cm]{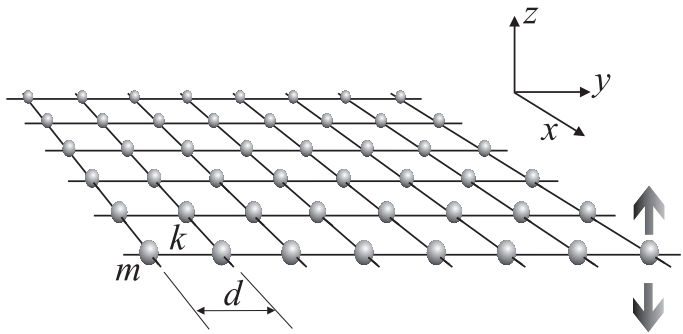}
\caption{\label{f1} Particle array representing the physical model
of the simulation.}
\end{figure}

The simulation developed in this work is based on the method
presented in \cite{thirty} and reproduces the propagation of
transverse mechanical waves. The correspondence between photonics
and the behavior of mechanical waves have been already reported in
the literature \cite{thirty-one}. The physical model consists in a
two-dimensional array of $p \times q$ particles of mass $m$
contained in the $x-y$ plane. Each particle is joined to its four
neighbors by means of elastic springs of elastic constant $k$ and
separated by a distance $d$. The movement of the particles is
constrained to the $z$-axis, which is normal to the plane
of the two-dimensional array, as shown in Fig.
\ref{f1}. The net force on each particle is null when it is
located at $z = 0$. A wave (that will propagate along the $x-y$
plane) is generated by applying an external force along the
$z$-axis to certain particles. The dynamic evolution of the system
along time is obtained by means of an iterative algorithm based on
an integration method of finite timesteps \cite{thirty}.

For large $p$ and $q$, the array of particles can be considered as a
continuous medium representing an elastic membrane subjected to a
tension per unit length $T = F_{e}/l_{T}$, where $F_{e}$ is the
elastic force and $l_{T}$  is the one-dimensional transversal
section. The simulated membrane has a surface mass density
$\mu=mN_{m}/s$, where $m$ stands for the mass of the mesh element
and $N_{m}$ is the number of particles contained
in a region of area $s$.  For a medium with ground density mass $\mu_{0}$,
the speed of the waves is given by $v_{0}=\sqrt{T/\mu_{0}}$. Therefore,
in a region with a mass density $\mu>\mu_{0}$ the speed of the waves is
$v<v_{0}$.
Making an analogy with optics, regions with mass density $\mu_{0}$ can be identified with
vacuum, i.e., a medium of refraction index $n_{0}=1$, while a region with an arbitrary mass density
$\mu$ corresponds to a medium with a real part of the refraction index $n=\sqrt{\mu/\mu_{0}}$.
This
approach permits to reproduce optical phenomena involving dielectric materials illuminated by
transverse electric (TE) polarized light in two-dimensional configurations, and this has been
verified for different systems and incidence configurations.

One of the remarkable features of this method is the
possibility of defining the simulation domain by means of digital
images or bitmaps. Each pixel in the image represents the
position of the particle in the array. In this
manner, a digital image of $p \times q$ pixels automatically
defines the size of the simulation domain. An appropriate
constant $\sigma_{p}$ given in [nm/pixel] links the size of an
object in the
digital image, in pixels, with the actual physical size of the sample to
be simulated, which is measured in nanometers.

Three bitmaps of equal size are defined within the simulation code to introduce different
characteristics of the structure and the illumination conditions, which are denoted as
$M$ (mass density), $D$ (damping) and $E$ (excitation).
The grey levels in the $M$ bitmap represent the mass
distribution within the array, which should be assigned using an adequate linear conversion
function.
Bitmap $D$ encodes the
damping constant of each point of the array, which allows introducing an
attenuation constant in the medium, and bitmap $E$ is introduced to
specify the excitation, i.e., the particles on which the external force is applied.
Following the analogy with
the propagation of electromagnetic waves in an optical medium, the $M$
bitmap determines the refraction index distribution in space,
bitmap $D$ specifies the regions where there is absorption, and
bitmap $E$ specifies the location and shape of light
sources. The grey level matrices $M$, $D$ and $E$
are related to the matrices
$M_{phys}$, $D_{phys}$ and $E_{phys}$, respectively, which
contain the values of mass, damping and excitation measured in
physical units, by:
\begin{equation}
M_{phys}= m_{0} + M m_{p}, \label{Eq00a}
\end{equation}
\begin{equation}
D_{phys}= D \mu_{p}, \label{Eq00b}
\end{equation}
\begin{equation}
E_{phys}= r_{p}\;(E-128), \label{Eq00c}
\end{equation}
where $m_{0}$ is the minimum mass value, which represents the
refraction index of vacuum. The proportionality constant $m_{p}$
in (\ref{Eq00a}) has units of [kg/gl] and the constant $\mu_{p}$
has units of [N sec/(m * gl)]. Notice that [gl] is the grey
level unit within a scale of 0-255 grey levels in which the
digital images are represented. $r_{p}$ in eq. (\ref{Eq00c}) is a
constant of units of [N/gl], which transforms the value of
grey level provided by the bitmap $E$ to a value of force. In eq.
(\ref{Eq00c}), a value $E=128$ indicates that no force is
applied on the particle. Then,
values over 128 are interpreted as positive forces and grey levels
values under 128 are interpreted as negative forces
\cite{thirty}. The harmonic excitation is introduced as
\begin{equation}
E_{t} = E_{phys} \sin{ ( \omega \,\tau_{n}\, n_{c} + \varphi)},
\label{Eq00d}
\end{equation}
where $E_{t}$ is the applied external force, $\omega$ is the
angular frequency of the excitation, $\varphi$ is the initial
phase and $\tau_{n}$ is a time adapting constant in units of
[sec/cycle] that converts the integer number of iteration cycles
$n_{c}$ --which define the timestep--, into a physical time variable.
The product $\tau_{n}\, n_{c}$ represents the discretized
time variable, and the product
$\omega_{d}=\omega\,\tau_{n}$ in eq. (\ref{Eq00d}) can be regarded
as a digitized angular frequency measured in [rad/cycle].

The Nyquist-Shannon sampling theorem \cite{NyquistShannon}
states that the frequency of the signal sampling must be
at least twice the highest frequency component of the
signal, in order to preserve the alternating nature of the external
excitation after the sampling. In our case, the sinusoidal
waveform of the externally applied force has a period $2\pi$, and
then it should be sampled as minimum at twice its frequency, that
is, every $\pi$ radians or less every iteration cycle.
This implies that $\omega_{d}$ should be smaller than $\pi$ rad/cycle.
On the other hand, the adapting constants $\tau_{n}$ and
$\sigma_{p}$ are related by
\begin{equation}
\tau_{n}=\frac{v_{d}}{v_{phys}} \sigma_{p},
\label{Eq00e}
\end{equation}
where $v_{d}$ is the digitized speed of the waves (measured in
[pixels/cycle]) and $v_{phys}$ is the physical speed of the waves
measured in [nm/sec]. In the case of optical waves in vacuum,
$v_{phys}=v_{0}$ corresponds to $c=2.99792 \times 10^{17}$
nm/sec. Therefore, the digitized angular frequency $\omega_{d}$
can be expressed as:
\begin{equation}
\omega_{d}=\omega \, \frac{v_{d} \, \sigma_{p}}{c}.
\label{Eq00f}
\end{equation}
The above expression implies that once the optical frequency
$\omega$ (or the optical wavelength $\lambda$) is
fixed, the Nyquist-Shannon criterion requires
\begin{equation}
v_{d}\, \sigma_{p}<\frac{\pi c}{\omega}.
\label{Eq00g}
\end{equation}

On the other hand, the
Courant-Friedrichs-Lewy condition \cite{thirty-two} imposes
that $v_{d0}$, the digital counterpart of the maximum allowed speed
$v_{0}$, must satisfy
$v_{d0} \leq 1$ pixel/cycle. That is, the maximum allowed
digitized wave speed that guarantees the stability of the
simulation is one pixel per cycle of iteration. A speed
beyond this value would cause the simulation to diverge. In other
words, the dynamical information can be transferred
to a maximum distance of one pixel, i.e., from one pixel
to the next one, in each iteration cycle. This limit is not related to the computer
processor speed or the physics of the modelled system. Instead, it
ensures an internal logical consistency of the numerical method.

The proposed method allows studying the response of any
two-dimensional distribution of refraction index in a very easy
and practical manner. The refraction index distribution can be
artificially generated using any available computational design
tool for digital image edition, or it can be obtained from a
digitized electron microscope image of a real physical structure.
Recently, Kolle {\em et al.} implemented an interface for the MEEP
package (an FDTD implementation) \cite{MEEP},
which permits introducing the profile of the diffracting structure
via binary images based on refraction
index contrasts in SEM or TEM images \cite{Kolle}.

Unlike other numerical methods in which the mathematical
expression of the problem is known and the computer is used as a
numerical integrator to obtain the solution, the idea behind the
proposed simulation method is the use of the computer as a
generator of a virtual environment where the physical differential
law is
used to make the system evolve along the time, as it would evolve
in the real physical world. The final solution is not predetermined by
any mathematical expressions, but by the configuration and causality of
the natural evolution of the introduced physical law.
This way of thinking the problem has the advantage of
naturally reproducing all the phenomena derived from
the characteristics of the simulated medium and from the physical law
involved. In this case, the simulated physical system
is able to reproduce the whole family of linear wave phenomena, such as
refraction, diffraction and interference, without having explicitly introduced
any wave equation and/or boundary conditions \cite{thirty}.

\section{Direction filter}
\label{sec03}

As mentioned above, the proposed simulation has the advantage of
genuinely reproducing the phenomena that would appear in a real
physical system. However, at the same time, this advantage is the
source of the main drawback of the method: its uncontrollability.
For instance, spurious reflected waves will naturally appear at
the edges of the simulation space. Incident and reflected waves
will be superimposed, and this superposition makes it impossible
to distinguish the amount of energy carried out by each one of the
waves. The energy density then results from the interference of
the incident and reflected waves. Besides, as it occurs in any
experimental setup, within the simulation is not possible to
determine the normal reflectance of a given sample under normally
incident light, because the detector cannot be placed in the path
of the incident light beam. And any artificial perturbation
introduced in the movement of the particles within the mesh in
order to decouple both waves, would affect the original (and
actual) response of the system.

Fortunately, in contrast to what happens in the physical world,
all the dynamical information of the system is known at every time
in the simulation. This allows extracting this information in
order to be processed in different ways. In the general case,
multiple waves will be travelling simultaneously in different
directions, producing a complex movement, and one would like
to decouple these different components from the complex collective
movement of the particles within the mesh.
A method used to isolate a wave
travelling in a given direction should be based on a
natural way of processing the information in order to be able to
work automatically for any wavelength, amplitude, direction,
waveform and, also, in the presence of an unknown number of
other waves travelling in different directions.

At a first sight, this does not seem a trivial task and therefore
the essential concept that represents a wave must be identified. A
travelling wave of any wavelength, amplitude, direction or
waveform has the following property: within a small region of
space (small enough to be considered homogeneous), the waveform
does not change in time, and only its phase changes at a known
rate given by its phase velocity. In this context, we present a
very simple and natural \emph{direction filter} (\emph{DF}) that
allows extracting a wave travelling in a given direction and with
a given sense. The filter is based on a wave subtraction technique
and only uses as input parameter the speed of the wave to be
filtered.

In order to present the direction filter in a clear
way, in the following subsection we describe its formulation for
one-dimensional problems and then, we present its extension to
more dimensions.

\subsection{Mathematical formulation of the \emph{DF} in 1D}
\label{sec03a}

\begin{figure}
\centering
\includegraphics[width=9cm]{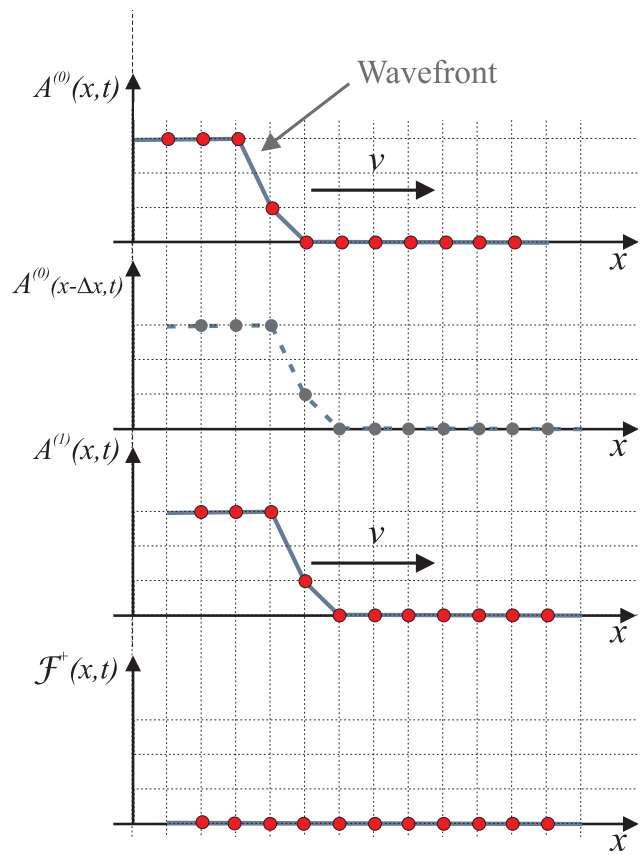}
\caption{\label{f2} Geometric representation of the action
performed by the \emph{positive DF operator}.}
\end{figure}

Suppose we have certain time-evolving function $A(x,t)$. Now, we
propose the following operator:
\begin{equation}
\mathcal{F}^{(+)}[A(x,t)]=A(x,t+\Delta t)-A(x-v\Delta t,t),
\label{Eq01}
\end{equation}
where $v$ is a constant representing a speed. If and only if
$A(x,t)$ is a wave travelling at speed $v$ towards the $+x$
direction, it must satisfy
\begin{equation}
A(x,t+\Delta t)=A(x-v\Delta t,t).
\label{Eq02}
\end{equation}
Substituting (\ref{Eq02}) in (\ref{Eq01}) we obtain
\begin{equation}
\mathcal{F}^{(+)}[A(x,t)]=A(x-v\Delta t,t)-A(x-v\Delta t,t)=0
\;\;\;\;\; \forall t.
\label{Eq03}
\end{equation}
$\mathcal{F}^{(+)}$ is called the \emph{positive DF
operator} acting on the wavefield $A(x,t)$ that cancels
waves travelling towards the $+x$ direction. Correspondingly,
$\mathcal{F}^{(-)}=A(x,t+\Delta t)-A(x+v\Delta t,t)$ is the
\emph{negative DF operator}, and it cancels waves travelling
towards the $-x$ direction. In other words, the $DF$ operator
cancels a wave $A(x,t)$ travelling in a given direction by
subtracting from it the same wave but evaluated in a previous
instant and in a position displaced by an amount $\Delta x=v
\Delta t$ from its present position. This is schematically shown
in Fig. \ref{f2}.

Let us now evaluate the effect of the \emph{positive DF
operator} in a more general case (the analysis for the
\emph{negative DF operator} is completely analogous). Suppose that
$A(x,t)=B^+(x,t)+B^-(x,t)$ is the superposition of two waves of
arbitrary shapes travelling with speed $v$ towards
opposite directions $+x$ and $-x$, respectively.
Taking into account that
\begin{equation}
A(x-v\Delta t,t)=B^+(x-v\Delta t,t)+B^-(x-v\Delta t,t)
\label{Eq04}
\end{equation}
and
\begin{equation}
A(x,t+\Delta t)=B^+(x,t+\Delta t)+B^-(x,t+\Delta t),
\label{Eq05}
\end{equation}
by applying $\mathcal{F}^{(+)}$ to this new
function, we get:
\begin{equation}
\mathcal{F}^{(+)}[A(x,t)]=B^+(x,t+\Delta t)+B^-(x,t+\Delta
t)-B^+(x-v\Delta t,t)-B^-(x-v\Delta t,t).
\label{Eq06}
\end{equation}
Since $B^+(x,t)$ and $B^-(x,t)$ are waves travelling towards the $+x$
and $-x$ directions, respectively, they satisfy
\begin{equation}
B^+(x,t+\Delta t)=B^+(x-v\Delta t,t)
\label{Eq07}
\end{equation}
and
\begin{equation}
B^-(x,t+\Delta t)=B^-(x+v\Delta t,t).
\label{Eq08}
\end{equation}
Therefore, the application of the \emph{positive DF
operator} results to be
\begin{equation}
\mathcal{F}^{(+)}[A(x,t)]=B^-(x+v\Delta t,t)-B^-(x-v\Delta t,t).
\label{Eq09}
\end{equation}
As expected, $B^+(x,t)$ is completely cancelled. By calling
$x'=x+v\Delta t$, eq.
(\ref{Eq09}) can be rewritten as
\begin{equation}
\mathcal{F}^{(+)}[A(x'-v\Delta t,t)]=B^-(x',t)-B^-(x',t-2\Delta t),
\label{Eq10}
\end{equation}
which represents a wave travelling towards the $-x$ direction. By
dividing both sides of eq. (\ref{Eq10}) by $2\Delta t$ and taking
the limit $\Delta t\rightarrow 0$, eq. (\ref{Eq10}) can be
expressed as:
\begin{equation}
\mathcal{F}^{(+)}[A(x,t)]\approx 2 \Delta t \frac{\partial
B^-(x+v\Delta t,t)}{\partial t}.
\label{Eq14}
\end{equation}
Expression (\ref{Eq14}) reveals the effect of the \emph{DF
operator} on a wave of general shape travelling in a non-filtered
direction. For small $\Delta t$ --compared with the time period of
the higher harmonic component of $B^-(x,t)$--, the
filtered wave is proportional to the time derivative of the
original wave.

\subsection{The \emph{Direction filter} in more dimensions}
\label{sec03b}

In general electromagnetic scattering problems we have
two- or three-dimensional wavefields, and then in this
subsection we generalize the \emph{DF} to more dimensions.

\begin{figure}
\centering
\includegraphics[width=9cm]{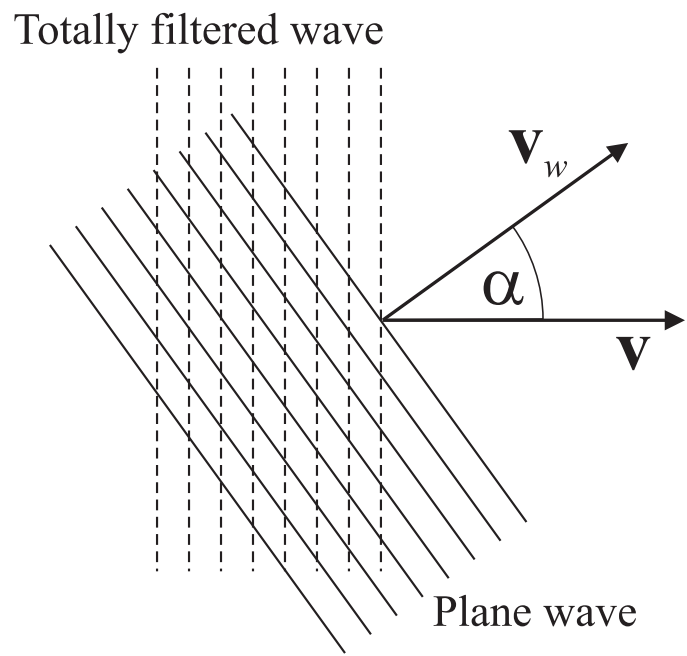}
\caption{\label{f3} Angle $\alpha$ between the direction of
propagation of the plane wave (along $\textbf{v}_{w}$) and the
filtering direction (along $\textbf{v}$).}
\end{figure}

Consider a scalar wavefield $A_{n}(\textbf{r},t)$ in a space of
$n$ dimensions. The general expression for the \emph{DF operator}
(\ref{Eq01}) in $\mathbb{R}^{n}$ is
\begin{equation}
\mathcal{F}_{n}^{(\textbf{v},\delta)}[A_{n}(\textbf{r},t)]=A_{n}(\textbf{r},t+\Delta
t)-A_{n}(\textbf{r}-\textbf{v}\delta,t),
\label{Eq17}
\end{equation}
This operator filters waves travelling with phase speed
$|\textbf{v}|$ in the direction of $\textbf{v}$, with a
characteristic time delay $\Delta t=\delta$. Note that bold
letters represent vectors. To find out the effect produced by the
\emph{DF operator} on waves travelling in directions different
from the filtering direction defined by $\textbf{v}$, we consider
a wavefield $A_{2}(\textbf{r},t)$ in $\mathbb{R}^{2}$, in which
case the \emph{DF operator} (\ref{Eq17}) becomes
\begin{equation}
\mathcal{F}_{2}^{(\textbf{v},\delta)}[A_{2}(\textbf{r},t)]=A_{2}(\textbf{r},t+\Delta
t)-A_{2}(\textbf{r}-\textbf{v}\delta,t).
\label{Eq19}
\end{equation}
If the wavefield is a two-dimensional plane wave
\begin{equation}
A_{2}(\textbf{r},t)=A\,e^{i\,\textbf{k}_{w}(\textbf{r}-\textbf{v}_{w}\,
t)}, \label{Eq18}
\end{equation}
where $\textbf{k}_{w}$ is the wave-vector, $\textbf{v}_{w}=\omega
/|\textbf{k}_{w}| \,\textbf{k}_{w}$ is the velocity of the
wave and $\omega$ its angular frequency,
substitution of (\ref{Eq18}) into (\ref{Eq19}) yields
\begin{equation}
|\mathcal{F}_{2}^{(\textbf{v},\delta)}(\alpha)|=A\, |e^{-i\,\omega\,
\delta}-e^{-i\,\omega \,\delta \,\cos(\alpha)}|,
\label{Eq20}
\end{equation}
where $|\mathcal{F}_{2}^{(\textbf{v},\delta)}(\alpha)|$ is the
complex amplitude of the filtered wave and $\alpha$ is the angle
between the propagation direction ($\textbf{k}_{w}$) and the
filtering direction ($\textbf{v}$) (see Fig. \ref{f3}).

To quantify the performance of the $DF$, we define the relative
attenuation $\mu_{a}$ of the wave as
\begin{equation}
\mu_{a}(\alpha)=1-|\mathcal{F}_{2}^{(\textbf{v},\delta)}(\alpha)|/|\mathcal{F}_{2}^{(\textbf{v},\delta)}|^{\rm max},
\label{Eq21}
\end{equation}
where $|\mathcal{F}_{2}^{(\textbf{v},\delta)}|^{\rm max}$ stands
for the maximum value of
$|\mathcal{F}_{2}^{(\textbf{v},\delta)}(\alpha)|$ for $\alpha \in
[0,360^\circ]$.

\begin{figure}
\centering
\includegraphics[width=11cm]{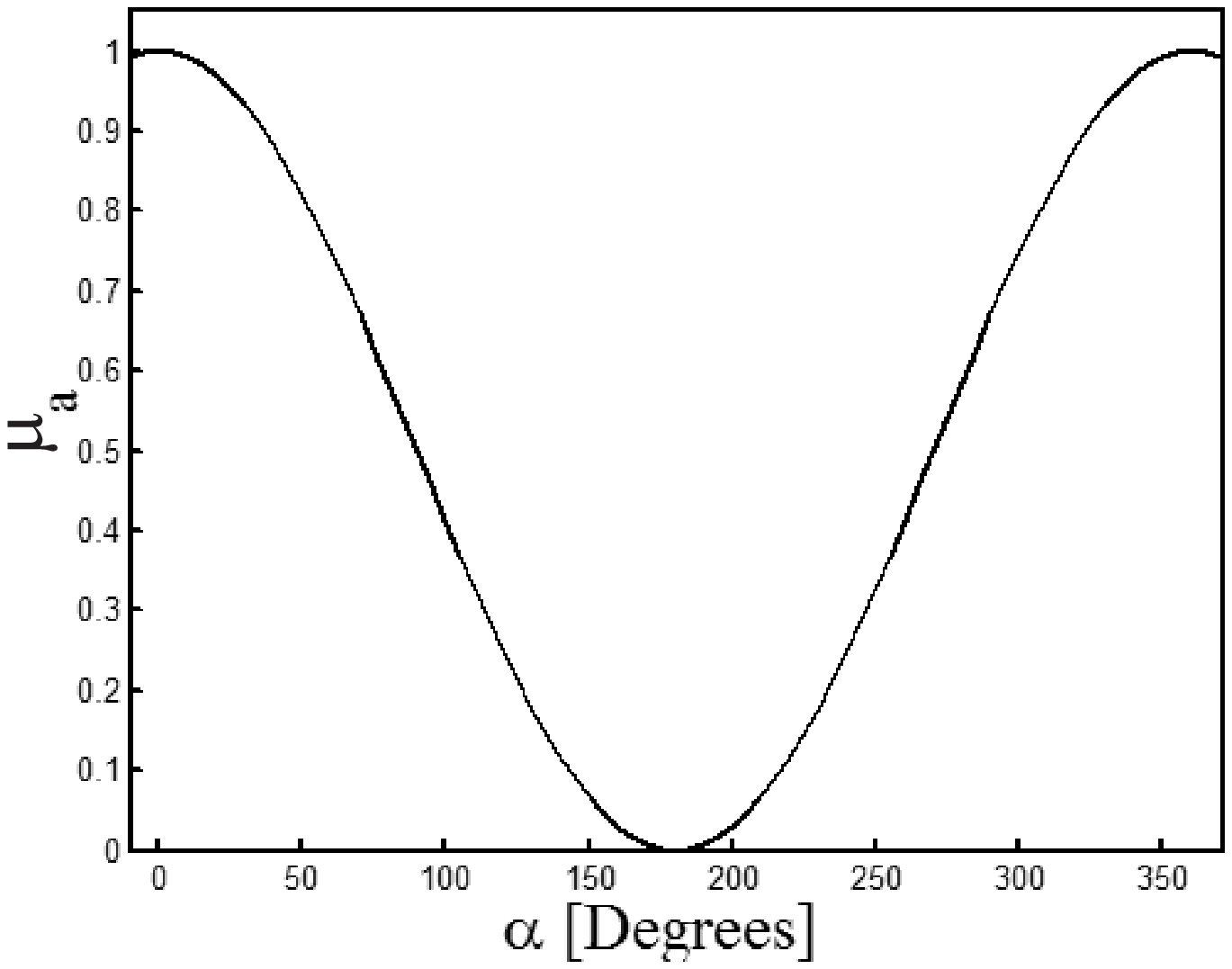}
\caption{\label{f4} Attenuation of the plane wave as a function of
the angle $\alpha$ between its direction of propagation and the
filtering direction.}
\end{figure}

\begin{figure}
\centering
\includegraphics[width=7cm]{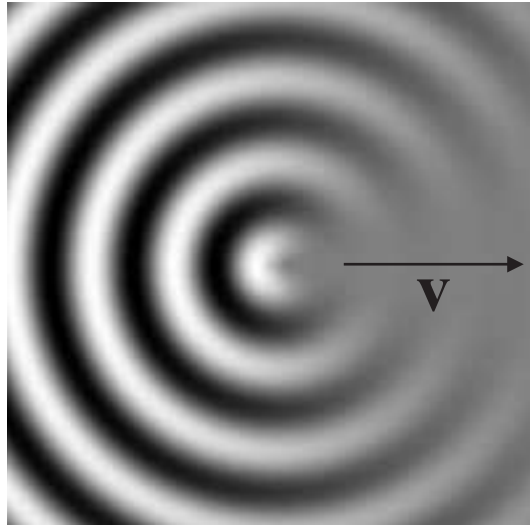}
\caption{\label{f5} Angular effect of the \emph{DF} on a circular
wavefront.}
\end{figure}

Figure \ref{f4} shows the relative attenuation of a plane wave as
a function of $\alpha$. As expected, the maximum attenuation is
obtained for $\alpha=0$, that is, when the propagation direction
of the plane wave coincides with the filtering direction.
Conversely, the attenuation is minimum for the wave travelling in
the opposite direction, that is, for $\alpha=180^\circ$. In Fig.
\ref{f5} we show the performance of the \emph{DF} for a circular
wavefront, which can be regarded as a superposition of plane waves
propagating along all directions. The amplitude of the obtained
filtered wave is shown in grey levels.

\subsection{\emph{Direction filter} implementation within the computer simulation}
\label{sec03c}

As presented above, the \emph{DF} is characterized by two parameters: $\textbf{v}$,
which determines the speed and direction of the wave to be filtered, and $\delta$,
the characteristic delay. As in any digital simulation, the space-time domain is
discretized and then, this discretization imposes certain constrains on the $DF$
parameters.

From eq. (\ref{Eq19}) it becomes evident that the filtering operation will
be effective if the spatially displaced
$A_{2}(\textbf{r}-\textbf{v}\delta,t)$ has exactly the same shape
as $A_{2}(\textbf{r},t+\Delta t)$. Since in the discretized domain
the minimum separation distance is 1 pixel, and the simulation
allows a maximum digitized wave speed $v_{d0}\leq 1$ pixel/cycle
(according to the Courant-Friedrichs-Lewy condition), the
travelling wave advances a distance $d_{0}=vt=v_{d0}$ $\times
\Delta n_{c}$ $\leq 1$ pixel in one iteration cycle ($\Delta
n_{c}=1$), and therefore, $A_{2}(\textbf{r},t+\Delta t)$ will have
the same shape as $A_{2}(\textbf{r}-\textbf{v}\delta,t)$ exactly
after $1/v_{d0}$ iteration cycles, with $(1/v_{d0})\in
\mathbb{Z}$. During intermediate iteration steps, the wave will
take interpolated values between
$A_{2}(\textbf{r}-\textbf{v}\delta,t)$ and
$A_{2}(\textbf{r},t+\Delta t)$ that will not exactly match the
shape of neither of them.

According to this, the necessary requirement for a good
performance of the \emph{DF} within the simulation is that the
speed of the waves must be set to $v_{d0}=1$ pixel/$\Delta n_{c}$,
with $\Delta  n_{c}\in \mathbb{Z}$ being the integer number of
iteration cycles in which the wave advances a distance of exactly
1 pixel. Therefore, the characteristic delay of the filter must be
$\delta=\Delta n_{c}=1$ pixel/$v_{d0}$ for a spatial shifting of
$A_{2}(\textbf{r}-\textbf{v}\delta,t)$ equal to 1 pixel. In our
simulations we usually set the second allowed digitized speed
$v_{d0}=0.5$ pixels/cycle, (i.e. $\Delta n_{c}=2$). Therefore, in
this case $\delta=2$ cycles for a spatial shifting of 1 pixel in
$A_{2}(\textbf{r}-\textbf{v}\delta,t)$. Although we could
eventually use spatial shiftings larger than 1 pixel, larger
shiftings lead to larger values of the characteristic delay
$\delta$, during which the shapes of
$A_{2}(\textbf{r}-\textbf{v}\delta,t)$ and
$A_{2}(\textbf{r},t+\Delta t)$ would be more affected by the
numerical errors, thus decreasing the quality of the directional
filtering.

It is worthwhile to mention that the above conditions are valid
for the implementation of the \emph{DF} in the orthogonal
directions $x$ and $y$, which coincide with the rows and columns
of the discretized array of particles comprising the simulation
medium, as shown in Fig. \ref{f1}. In the present paper we
use four \emph{DF}s corresponding to the directions $+x$, $-x$,
$+y$ and $-y$.
From a practical programming point of view, the \emph{DF} is applied to the waves evolving in the
two-dimensional array defined by the simulation, called \emph{main plane}.
However, the resulting directionally filtered
waves are stored in \emph{secondary planes}. These
\emph{secondary planes} show time-evolving waves that are
filtered images of the waves evolving in the \emph{main plane}, and
therefore, the \emph{DF} does not affect the physics of
the system. In this case, we have four secondary
planes that store the filtered waves travelling
in the $+x$, $-x$, $+y$ and $-y$ directions, respectively.

\begin{figure}
\centering
\includegraphics[width=12cm]{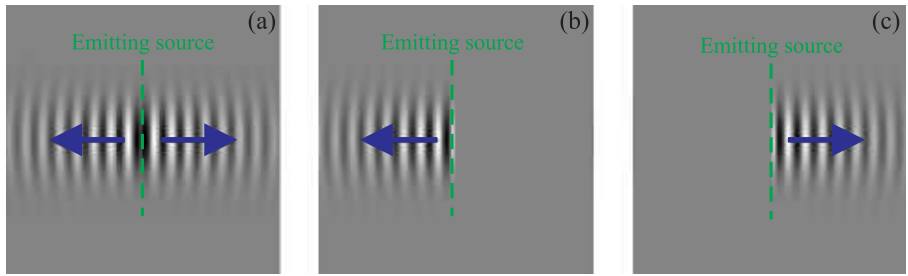}
\caption{\label{f6} (a) Source emitting gaussian waves in
directions $+x$ and $-x$. (b) \emph{Secondary plane} of the
\emph{DF} that cancels waves travelling towards the $+x$
direction. (c) \emph{Secondary plane} of the \emph{direction
filter} that cancels waves travelling towards the $-x$ direction.
Blue arrows show the propagation direction of the wavefronts.}
\end{figure}

To illustrate the behavior of the $DF$, in Fig. \ref{f6}(a) we show
the \emph{main plane} of the simulation of a linear source emitting
gaussian waves in both directions $\pm x$. Figures \ref{f6}(b) and
\ref{f6}(c) show the \emph{secondary planes} that contain the
wavefield of Fig. \ref{f6}(a) after the application of the
\emph{DF} that cancels waves travelling towards the $+x$ and the
$-x$ direction, respectively.

\subsection{Determination of the energy flux vector field}
\label{sec03d}

One of the main applications of
the \emph{DF} presented in the previous subsections is the method
for evaluating the energy flux of any simulated wavefield.
This method makes use of the waves filtered in the orthogonal directions
$\pm x$ and $\pm y$.
Let us call $I_{x+}$, $I_{x-}$, $I_{y+}$ and $I_{y-}$ the intensity
(calculated as the time integration of the square of the amplitude) of the
waves travelling towards the $+x$, $-x$, $+y$ and $-y$ directions,
respectively. Then, the quantities
\begin{equation}
\phi_{x}=I_{x+}-I_{x-} \label{Eq22}
\end{equation}
and
\begin{equation}
\phi_{y}=I_{y+}-I_{y-} \label{Eq23}
\end{equation}
are proportional to the energy flux along the $x$ and $y$
directions, respectively.
Consequently, the vector field $\mathbf{F}$ defined as
\begin{equation}
\mathbf{F}=(\phi_{x},\phi_{y}) \label{Eq25}
\end{equation}
is proportional to the energy flux vector field.

According to (\ref{Eq14}), the
filtered intensities are not equal to the intensity
of the original wavefield, but proportional to its derivative.
However, to determine
the direction of the energy flux --and not its absolute
value--, it is enough to compute the relative intensities between
the filtered waves.

As an example, let us calculate $\phi_{x}$ for a wave $Z(\textbf{r},t)$ travelling
along the $+y$ or the $-y$ direction. Since
\begin{equation}
\mathcal{F}_{2}^{(+x,\delta)}[Z(\textbf{r},t)]=\mathcal{F}_{2}^{(-x,\delta)}[Z(\textbf{r},t)],
\label{Eq24}
\end{equation}
i.e., the \emph{DF}s for the $+x$ and $-x$ directions
produce the same attenuation on waves that propagate along a
direction orthogonal to their corresponding directions of filtering (see eq. (\ref{Eq20})),
according to (\ref{Eq22}) $\phi_{x}$ becomes
\begin{equation}
\phi_{x}=I_{x+}-I_{x-}=\int(\mathcal{F}_{2}^{(+x,\delta)})^2
dt-\int(\mathcal{F}_{2}^{(-x,\delta)})^2 dt=0. \label{Eq26}
\end{equation}
As expected, there is no energy flux along the
$x$ direction for the evaluated wave.

It should be mentioned that the \emph{DF} developed here could
also be applied to experimental data. For instance, the method of
\emph{Fourier transform profilometry} presented in
\cite{thirty-four} permits digitalizing the evolution of surface
water waves along time. Therefore, the technique proposed in this
paper also enables the calculation of the energy flux field for real
systems.

\section{Dynamic differential absorber}
\label{sec04}

It is well known that any computer
wave simulation can only reproduce the propagation of waves in a
finite domain. Therefore, the waves arriving to the edges of the
simulation domain will be naturally reflected back \cite{thirty}.
In order to simulate open boundaries, it is necessary to artificially cancel
all the waves reflected at the edges of the simulation
space. For this purpose, several methods have been proposed
\cite{thirty,thirty-five,thirty-six}, each of them having its
advantages and disadvantages.
One possibility to avoid these reflections is to place a slab of
an absorbing medium adjacent to the edges of the domain.
In its basic implementation, this approach reduces the reflected waves,
but it does not
completely cancel them \cite{thirty}. Besides, the absorbing region produces an
unnecessary increment of the size of
the simulation space and, consequently, the computation time also increases.

A highly improved version of the procedure described above is the
so called \emph{perfectly matched layer} (PML) method
\cite{twenty-seven,thirty-six}. This method is the most widely
spread technique to cancel reflected waves, and it consists in
introducing an absorbing anisotropic layer at the edges of the
simulation space. Within this layer, the differential wave
equation is modified by including a special transformation that
produces a rapid attenuation of the wave as it propagates.
Although this method results to be very efficient, it requires a
layer of finite thickness to allow the decay of the waves.
Besides, waves travelling parallel to the layer are not attenuated
by the PML method, producing the accumulation of non-realistic
energy in that region.

In this Section we present an alternative approach to simulate the
open space by cancelling
reflected waves at the edges of the simulation domain. The method,
called \emph{dynamic differential absorber} (DDA), is
based on an intuitive concept, and its main advantage is that it does not
require a layer of a given thickness to cancel the waves, i.e., it produces
the absorption of the wave within a layer of infinitesimal thickness,
and this saves computation space and time.
On top of that, the method automatically cancels waves of any
amplitude, shape or frequency with the same efficiency, and this constitutes a
great advantage that enables the use of multi-frequency excitation, as
shown in Section 5. As in the case of the
\emph{DF}, only the velocity of the incoming waves should be specified.
It is important to mention that this method can be applied to waves
of any nature, i.e., mechanical, electromagnetic, and potentially
even to quantum waves governed by Schrodinger's equation.

In the following subsection we present the basic idea by means of the 1D
formulation of the DDA, called here the \emph{simple dynamic
differential absorber} (SDDA). Then, we develop the
\emph{adaptive dynamic differential absorber} (ADDA), which
is a generalization to two or more dimensions, and in which case the
direction of the incoming waves must be taken into account.

\subsection{Simple dynamic differential absorber (SDDA)}
\label{sec04a}

\begin{figure}
\centering
\includegraphics[width=12cm]{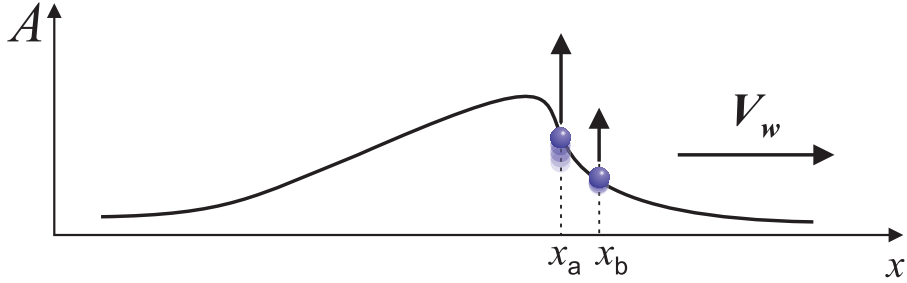}
\caption{\label{f7} Amplitude of a propagating wave
at two fixed points located in $x=x_{a}$ and $x=x_{b}$.}
\end{figure}

Consider a wave propagating along the \emph{x}-axis towards the
$+x$ direction, and let us examine the movement of this wave at
two fixed points $x=x_{a}$ and $x=x_{b}$, as shown with the blue
dots in Fig. \ref{f7}. As the wave passes through, the amplitude
$A_{b}$ at $x_{b}$ will take the value of $A_{a}$ at $x_{a}$, but
with a time delay given by $\Delta t= \Delta d / v_{w}$, where
$\Delta d=x_{b}-x_{a}$ and $v_{w}$ is the phase velocity of the
wave. Within this context, the term ``amplitude'' refers to the
magnitude of the wavefield at a given point.

The key idea is the following: if we artificially control the
amplitude $A_{b}$ of the wave at $x_{b}$ and make it move in such
a way that $A_{b}$ copies the amplitude $A_{a}$ with the right
time delay, the behavior of the propagating wave for $x<x_{b}$
will not be affected as it passes through. In fact, if $x_{b}$ is
a point located at the edge of the simulation domain, the incoming
wave will not be affected at this point and it would continue
propagating as if there were no boundary. Therefore, the SDDA
consists in controlling the movement of a point located at
$x_{b}$, called the \emph{absorbing} point, according to the
behavior of the incoming wave at $x_{a}$, called the
\emph{reading} point. Mathematically, we can exprees this
procedure as
\begin{equation}
A_{b}^{(t_{1})}=A_{a}^{(t_{0})}, \label{Eq27}
\end{equation}
where $A_{b}^{(t_{1})}$ is the amplitude of the wave at $x_{b}$ in
a certain instant $t_{1}$, $A_{a}^{(t_{0})}$ is the amplitude of
the wave at $x_{a}$ in a previous instant $t_{0}=t_{1}-\Delta t$,
where $\Delta t=\delta_{d}\; \Delta d$.
In this context, $\delta_{d} = 1/v_{w}$ is the \emph{differential
delay} which depends on the medium
characteristics via the phase velocity of the waves within the
propagating medium.

Although the above equations indicate that the \emph{reading}
point can be located arbitrarily close to the \emph{absorbing}
point, within the simulation method the discretized nature of the
space-time domain must be taken into account, as already explained
in subsection \ref{sec03c} for the \emph{DF}. Since the minimum
distance between the \emph{reading} and the \emph{absorbing}
points is of one pixel, we also set the allowed digitized wave
speed $v_{d0}=0.5$ pixels/cycle (see subsection \ref{sec03c}), and
then $\delta_{d} = 1/v_{d0}=2$ iteration cycles. If, for instance,
a 1D simulation space is $M$ pixels long, the rightmost pixel
$m=M$ is forced to move as:
\begin{equation}
A_{M}^{(n_{c})} = A_{M-1}^{(n_{c}-2)},
\label{Eq30}
\end{equation}
where $A_{M-1}^{(n_{c}-2)}$ is the amplitude of the wave at the
pixel $M-1$ stored two iteration cycles before the present cycle
$n_{c}$. Similarly, the leftmost pixel $m'=1$ is forced to move
as:
\begin{equation}
A_{1}^{(n_{c})} = A_{2}^{(n_{c}-2)},
\label{Eq31}
\end{equation}
where $A_{2}^{(n_{c}-2)}$ is the amplitude of the second pixel
stored two iteration cycles before the present cycle $n_{c}$.

\subsection{Adaptive dynamic differential absorber (ADDA)}
\label{sec04b}

If the simulation domain is two-dimensional, the SDDA presented in the previous
subsection would only be effective for waves propagating
normally to the edge of the simulation space (or whose
wavefronts are parallel to the edges of the domain). In order to
develop a direction-sensitive method that could properly absorb
waves propagating with different directions,
small corrections must be introduced into the \emph{differential delay}
used to cancel the incoming waves.

\begin{figure}
\centering
\includegraphics[width=9cm]{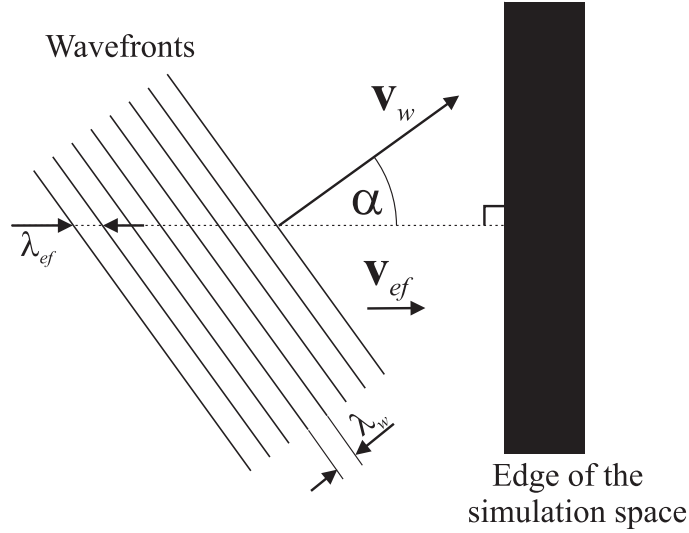}
\caption{\label{f8} Effective wavelength
$\lambda_{\rm eff}$ and effective phase velocity $v_{\rm eff}$ of an
obliquely incident wave forming an angle $\alpha$ with the normal
to the simulation space edge.}
\end{figure}

The effective wavelength $\lambda_{\rm eff}$ of the wave that arrives at
the edge of the simulation space is given by
\begin{equation}
\lambda_{\rm eff} = \lambda_{w}/ \cos(\alpha), \label{Eq32}
\end{equation}
where  $\lambda_{w}$ is the actual wavelength of the incoming wave and
$\alpha$ is the angle between the direction of propagation
and the normal to the edge of the simulation space.
Then, the effective phase velocity $v_{\rm eff}$ --the phase velocity
of the wave along the direction normal to the edge of the
simulation space--, depends on the angle of incidence and is given by:
\begin{equation}
v_{\rm eff}=\frac{\omega}{k_{\rm eff}}=\lambda_{\rm eff} f,
\label{Eq33}
\end{equation}
with $\omega= 2\pi f$, $k_{\rm eff}$ being the effective wavenumber
and $f$ the frequency of the wave. This is schematically shown in
Fig. \ref{f8}. Consequently, the \emph{effective
differential delay} $\delta_{\rm eff}$ becomes
\begin{equation}
\delta_{\rm eff} = \frac{1}{v_{\rm eff}}=\frac{\cos(\alpha)}{v_{w}}=\delta_{d} \cos(\alpha),
\label{Eq34}
\end{equation}
which implies that $\delta_{\rm eff}$ is always
smaller than $\delta_{d}$.

\begin{figure}
\centering
\includegraphics[width=11cm]{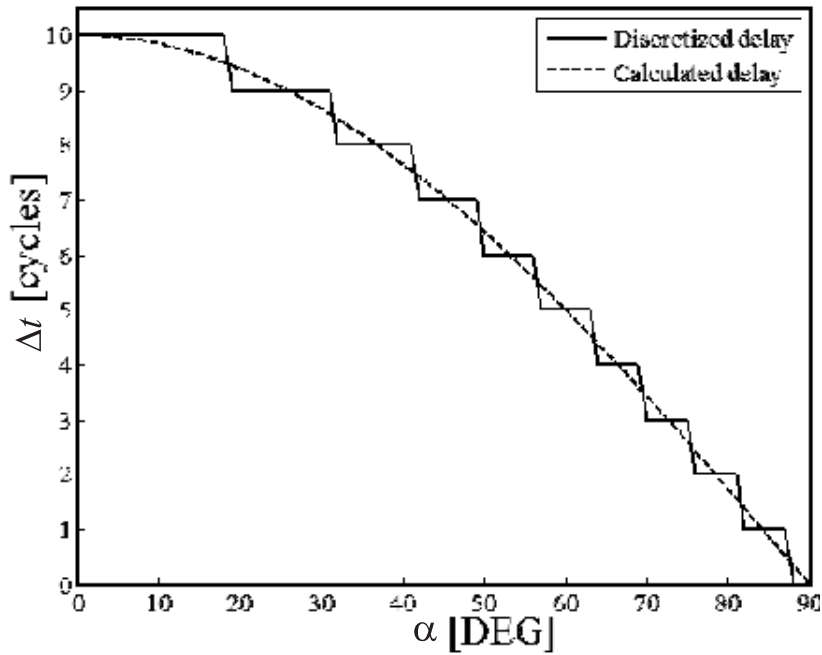}
\caption{\label{f9} Delay vs. angle of incidence: discretized
(solid line) and calculated (dashed line).}
\end{figure}

\begin{figure}
\centering
\includegraphics[width=9cm]{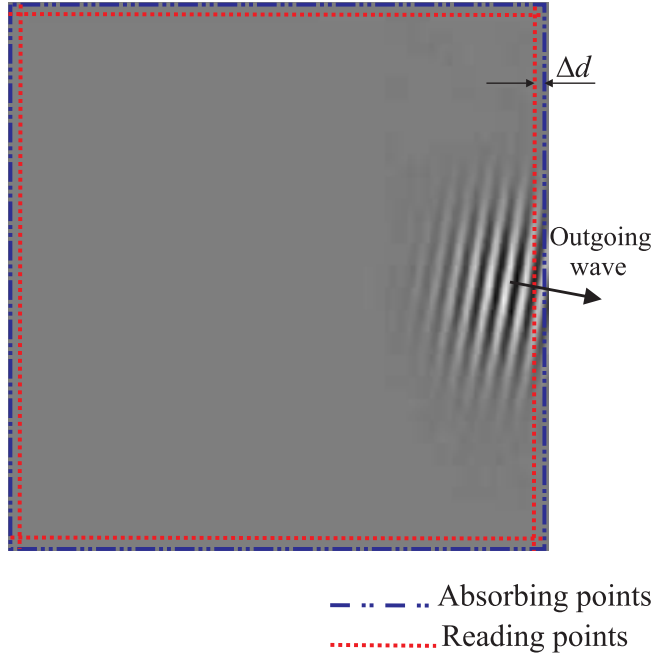}
\caption{\label{f10} Location of the \emph{absorbing} and the \emph{reading}
points for a 2D simulation space. The distance $\Delta d$ is also
indicated.}
\end{figure}

To determine $\delta_{\rm eff}$, the \emph{reading} point must provide
information not only about the time-varying amplitude of the wave
to be cancelled, but also about its direction of propagation,
given by $\alpha$. This angle can be obtained by evaluating
the energy flux vector at the \emph{reading} point, using the method
described in subsection \ref{sec03d}. For example, for the
right hand-side edge of the simulation space, $\alpha$
is given by
\begin{equation}
\alpha = \arctan (\frac{\phi_{y}}{\phi_{x}}), \label{Eq36}
\end{equation}
where $\phi_{x}$ and $\phi_{y}$ are the $x-$ and $y-$components of
the energy flux, defined in eqs. (\ref{Eq22}) and
(\ref{Eq23}).
In practice, to clearly establish the direction of the
incoming wave before it is affected by the edge of the simulation
space, the information about the direction of the incoming waves
is taken from points located between 2 and 4 pixels away from the
edge of the simulation space, where the absorbing points are
located. On the other hand, the
discretized nature of the simulation must be taken into account.
If the wave speed is set to $v_{d0}=0.5$
pixels/cycle, $\delta_{\rm eff}$ is a real number between 2 and 0
for $\alpha$ between 0
and 90$^\circ$, respectively. Since the number of cycles $\delta_{\rm eff}$ must be
an integer, the resulting value $\Delta t$ must be approximated to
the closest integer value. This implies that the allowed
discretized delays ($\delta_{\rm eff}=$ 2, 1 and 0) will only match
the required delay for three angles of incidence, that
in this case are $\alpha=$ 0$^\circ$, 60$^\circ$ and 90$^\circ$.
Therefore, in order to increase the number of allowed discretized
delays, the wave speed can be decreased, and set, for instance, to
$v_{d0}=0.1$ pixels/cycle. In this case, $\delta_{\rm eff}$ will be
bounded between 10 and 0, for $\alpha$ between 0$^\circ$ and 90$^\circ$,
respectively. Then, the number of discretized delays is increased,
and the error between the calculated and the allowed integer
delays is minimized, as shown in Fig. \ref{f9}. Increasing the
number of discretized delays improves the effectiveness of the
ADDA. Then, to increase the number of discrete delays we can
either increase the distance between the \emph{reading} and the \emph{absorbing}
point, or decrease the wave speed $v_{d0}$. Depending on the
requirements of speed and usable space size of the simulation, the
best choice should be made. Figure \ref{f10} shows the location of
the \emph{absorbing} points (in blue) and the \emph{reading} points (in red) for
a two-dimensional space. The distance $\Delta d$ (magnified for
clarity) between the \emph{reading} and the \emph{absorbing} points is also
indicated.

Despite the difference between the calculated and the discrete
values of $\delta_{d}$ introduced for certain incidence angles by
the discretization process, it was verified that these errors do
not significantly affect the performance of the ADDA, even for
$\delta_{d}=2$ (second maximum digitized wave speed allowed) and
$\Delta d=1$ (maximum available space), as will be shown in the
next subsection. Making an analogy with electronics, the proposed
method behaves like an active device, which reads an input signal
and reacts accordingly to return a post-processed output signal.
In this sense, this method is different from most available
absorbing methods, which behave like passive devices whose
response does not take into account the input signal.

\subsection{Validation}
\label{sec04c}

In order to evaluate the performance of the ADDA, the reflectance,
defined as the ratio of the reflected to the incident intensity,
was calculated for a gaussian beam incident on the edges of the
simulation space with different angles.

\begin{figure}
\centering
\includegraphics[width=11cm]{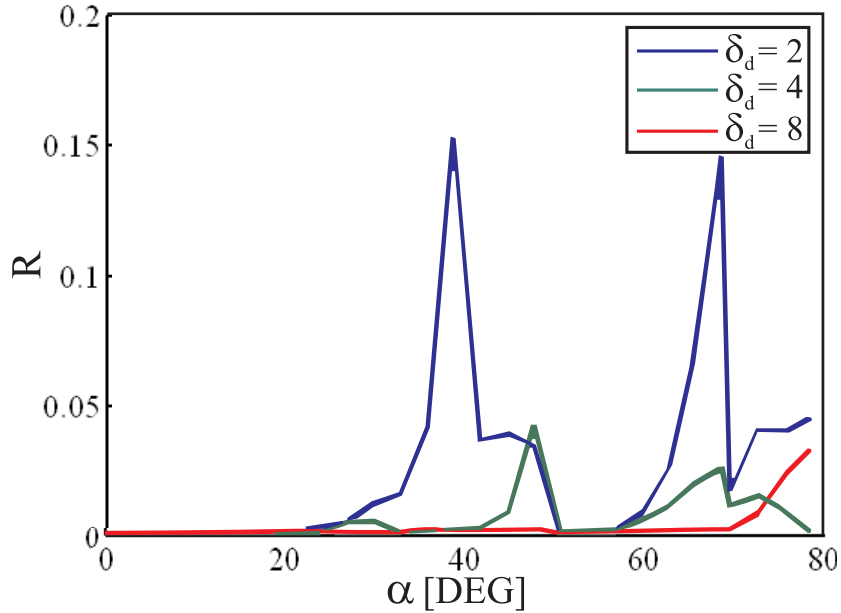}
\caption{\label{f11} Numerical experiment with the ADDA:
Reflectance vs. angle of incidence $\alpha$ for $\delta_{d}=2$
(blue line), $\delta_{d}=4$ (green line) and $\delta_{d}=8$ (red
line).}
\end{figure}

In all cases, the distance between the \emph{reading} and the
\emph{absorbing} points was set to $\Delta d=1$ pixel. Figure
\ref{f11} shows the reflectance as a function of the angle of
incidence for $\delta_{d}=2$ (blue line), $\delta_{d}=4$ (green
line) and $\delta_{d}=8$ (red line).
It can be observed that as $\delta_{d}$ is increased, the overall
reflectance decreases, that is, a better performance of the
absorber is obtained. For $\delta_{d}=2$, there are peaks of
relatively high reflectance for 39$^\circ$ and 69$^\circ$. In the
case of $\delta_{d}=4$ the reflectance also shows peaks, but less
intense than for $\delta_{d}=2$. These peaks are located at the
angles of incidence for which the difference between the
calculated and the available integer delays is maximized. As the
number of integer delays is increased (by increasing
$\delta_{d}$), a better matching between the calculated and the
available integer delays is obtained (see Fig. \ref{f9}), and the
intensity of the reflectance peaks is gradually reduced, as
observed for $\delta_{d}=8$ in Fig. \ref{f11}.

\begin{figure}
\centering
\includegraphics[width=11cm]{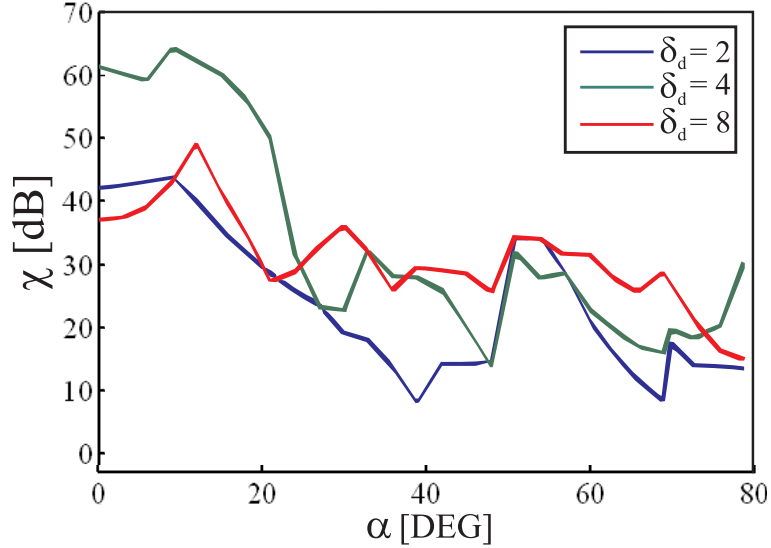}
\caption{\label{f12} Attenuation vs. angle of incidence $\alpha$
for $\delta_{d}=2$ (blue line), $\delta_{d}=4$ (green line) and
$\delta_{d}=8$ (red line).}
\end{figure}

Figure \ref{f12} shows the attenuation produced by the ADDA
expressed in decibels (dB), and calculated as $\chi =-10\,
\log_{10}(R)$, where $R$ is the reflectance. It can be noticed that
for the angles for which the matching between the integer
and the calculated delays is better, the attenuation values reach up to 44.2 dB for
$\delta_{d}=2$, 64.5 dB for $\delta_{d}=4$ and 49.4 dB for
$\delta_{d}=8$. As a reference, the maximum
attenuation values obtained with the PML applied within the FDTD
method lie between 20 and 39 dB for small incidence angles,
depending on the setting parameters of this method
\cite{thirty-six}.

It can also be observed in Fig. \ref{f12} that for angles of
incidence greater than approximately 25$^\circ$, the best (larger)
attenuation is obtained for $\delta_{d}=8$, whereas for small
angles of incidence the attenuation obtained for $\delta_{d}=4$ is
better. This result can be explained by taking into account that
smaller angles of incidence require higher values of the effective
differential delay $\delta_{\rm eff}$, as determined by
(\ref{Eq34}). In this case, there are more iteration cycles
between the \emph{reading} and the \emph{absorbing} instants for
$\delta_{d}=8$ than for $\delta_{d}=4$. Then, if the sampling
resolution of the shortest wavelength involved is low, a wave that
propagates a fraction of pixel will be represented in the next
timestep by a natural interpolation produced by the simulation.
When the wave advances one pixel, the accumulated numerical errors
produced by interpolation will be higher in the case of
$\delta_{d}=8$ than in the case of $\delta_{d}=4$, producing a
higher mismatch between the read waveform and the waveform that
actually reaches the absorbing point. This produces a better
performance for $\delta_{d}=4$ for small angles of incidence.
Therefore, in order to reduce the interpolation errors and improve
the absorbing characteristics for small incidence angles, --i.e.,
for high differential delays $\delta_{d}$,-- the sampling
resolution of the wavelengths comprising the wave to be cancelled
must be increased, and this is done by decreasing the constant
$\sigma_{p}$ described in Section \ref{sec02}.

\begin{figure}
\centering
\includegraphics[width=9cm]{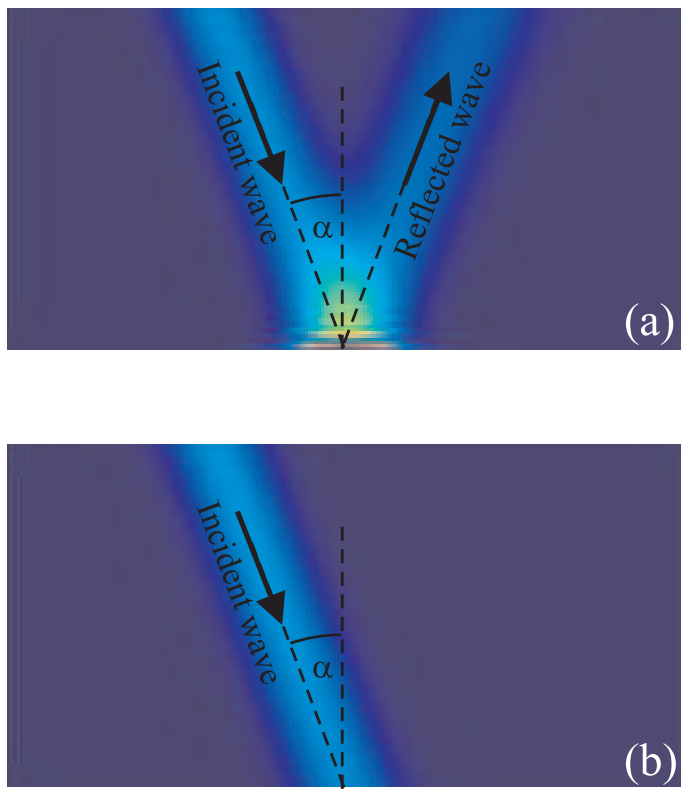}
\caption{\label{f13} Intensity diagram of a gaussian beam forming
an angle $\alpha=$ 20$^\circ$ with the normal to the lower
horizontal edge of the simulation space for the case
$\delta_{d}=4$. (a) without ADDA; (b) with ADDA.}
\end{figure}

The effect of the ADDA is graphically shown
in Fig. \ref{f13} for an incident gaussian beam forming an angle
of 20$^\circ$ with the normal to the edge of the simulation space.
Figure \ref{f13}(a) shows the intensity diagram without the ADDA,
where the reflected wave can be observed.
On the other hand, Fig. \ref{f13}(b) shows that the reflected beam
is almost completely cancelled when the absorber is activated.

\section{Tuning filter}
\label{sec05}

The optical response of a structure is usually described by its
reflectance and transmittance as a function of the wavelength.
Therefore, in order to obtain its optical response
using a simulation, the program should be executed for many
frequency values and this could take considerable computing time,
especially if the desired spectral resolution is high. To avoid
this problem, in this Section we show that a \emph{multi-frequency
excitation} (MFE) scheme can be implemented within the simulation
in a quite straightforward manner.

Within this framework, the single frequency excitation
(SFE) described in (\ref{Eq00d}) is replaced by \begin{equation}
E_{t} = \sum_{i=1}^{f_{\rm tot}} E_{i} \sin ( \omega_{i}\,\tau_{n}\,
n_{c} + \varphi_{i}), \label{EqSec05a}
\end{equation}
where $f_{\rm tot}$ is the total number of frequencies, $E_{i}$ is the
excitation bitmap for the $i$-th frequency $\omega_{i}$ and
$\varphi_{i}$ is the $i$-th initial phase. In what follows we set
$E_{i}=E$, meaning a uniform frequency spectrum.

\begin{figure}
\centering
\includegraphics[width=12cm]{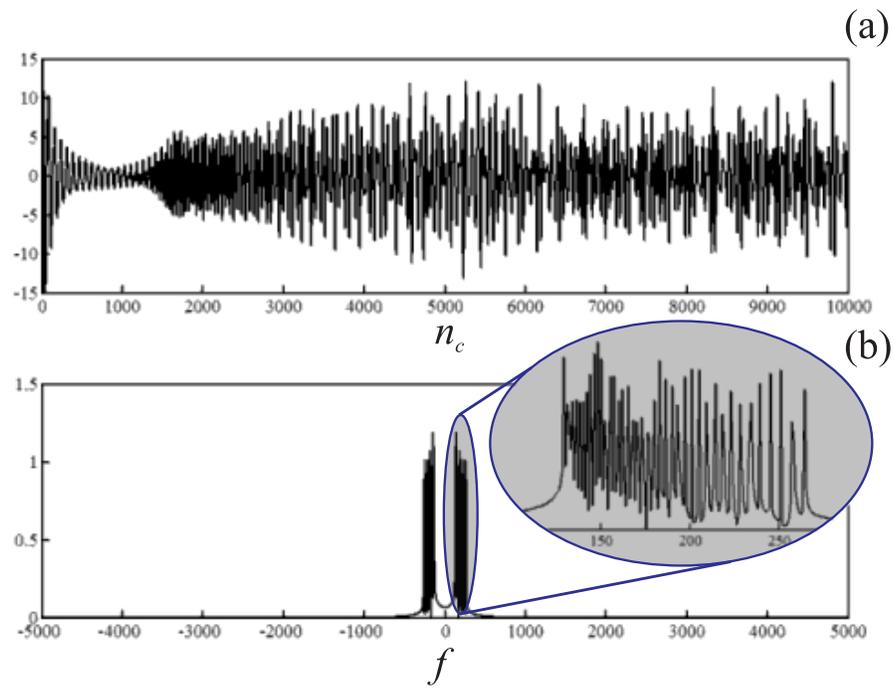}
\caption{\label{f14} Multi-frequency excitation corresponding to
41 optical wavelengths contained within the range 380 - 780 nm,
for $n_{c}^{\rm tot}=10000$ cycles, $\tau_{n}=3.335*10^{-17}$
sec/cycle and equal magnitude of the external force applied to all
the waves of different frequencies: (a) time evolution of a single pixel; (b) the
absolute value of its FFT.}
\end{figure}

Each point in the simulation space has a complex oscillating
movement resulting from a mixture of the whole set of waves of
different frequencies having unknown amplitudes, and we are
interested in extracting each frequency component from the
multi-frequency wavefield. One possible way to do this is to store
the time evolution of the whole set of pixels in the simulation
space in order to perform the Fourier transform for each pixel.
Then, the spectral amplitude for each pixel can be obtained from
the Fourier spectrum evaluated at any single frequency. The
maximum number of iteration cycles used in each particular case
depends on the size of the bitmap that defines the simulation
space and on the setting of the digitized speed of the wave
$v_{d0}$. In order to guarantee that the waves could reach every
point within the simulation space, we established the following
criterion: the total number of iteration cycles $n_{c}^{\rm tot}$
should be such that it ensures that a wave can travel twice the
longest straight distance within the simulation space. For a
bitmap of $p \times q$ pixels and a wave speed $v_{d0}$, the
minimum number of cycles is:
\begin{equation}
n_{c}^{\rm tot}=2\,\frac{\sqrt{p^{2}+q^{2}}}{v_{d0}}.
\label{EqSec05b}
\end{equation}
Therefore, to evaluate the Fast Fourier Transform (FFT) for each
pixel, it would be necessary to store the complete evolution of
the waves in a 3D matrix of size $(p,q,n_{c}^{\rm tot})$. Hence, for a
3D simulation space, a 4D matrix should be stored. For example, a
2D simulation space of 1 Mpixel would require to store a 21 GBytes
matrix (with double precision), which exceeds the RAM memory of
today conventional computers. On the other hand, in the FFT
algorithm the frequency domain size equals the digitized time
domain size. Then, for a typical resolution of 10 nm in the
wavelength, the set of points corresponding to the whole set of
frequencies results to be very localized in the Fourier space, as
shown in Fig. \ref{f14}. Under these conditions, and due to the
digitized nature of the Fourier space, the estimation of each
spectral amplitude is highly inaccurate. In order to reduce the
error, the size of the frequency domain should be enlarged by
increasing even more the size of the digitized time domain
$n_{c}^{\rm tot}$ given by eq. (\ref{EqSec05b}). Figure \ref{f14}(a)
shows the amplitude (in arbitrary units) of the time evolution of
a pixel under multi-frequency excitation for 41 simultaneous waves
of optical frequencies contained within the range 380 - 780 nm and
using $n_{c}^{\rm tot}=10000$ cycles, $\tau_{n}=3.335*10^{-17}$
sec/cycle and the same excitation amplitudes for all the
frequencies. The amplitudes of the detected frequencies are very
irregular, showing that the FFT is not useful to adequately
isolate the amplitude of each frequency under these excitation
conditions and digital resolution, as shown in the example of Fig.
\ref{f14}(b).

To solve this problem, we introduce the \emph{tuning filter}
(\emph{TF}), that allows extracting a single-frequency
time-varying wave from the multi-frequency wavefield. This filter
acts as a temporal mask which is applied on the simulation space.
After the application of the \emph{TF} to the time-varying
multi-frequency wavefield, a dynamic single-frequency wave can be
visualized, allowing its processing to determine physical
magnitudes such as reflectance or transmittance for a given
frequency.

\subsection{Implementation}
\label{sec05a}

The \emph{TF} is based on the tuning properties of the forced
damped oscillator, governed by the well known inhomogeneous
differential equation
\begin{equation}
\frac{F_{\rm ext}}{m}=\frac{d^{2}z}{dt^{2}}+\gamma \;\frac{dz}{dt} +
\omega_{do}^{2}\;z, \label{EqSec05c}
\end{equation}
where $F_{\rm ext}$ is the applied external force, $z$ and $t$ are the
position and time variables, respectively, $\gamma$ is the damping
constant, and $\omega_{do}=\sqrt{k_{do}/m_{do}}$ is the natural
frequency of the damped oscillator ($k_{do}$ is the spring elastic
constant and $m_{do}$ is its mass). The \emph{TF} uses the characteristic
frequency response of the forced damped oscillator, which acts as
a narrow band filter and maximizes the signal at resonance, i.e.,
when the frequency of $F_{\rm ext}$ equals the natural frequency $\omega_{do}$ of the
oscillator. Under the condition $\gamma \ll \omega_{do}$, the
width of the resonance peak is given by
\begin{equation}
\Delta \omega \approx \gamma
\label{EqSec05d},
\end{equation}
the resonant frequency is given by
\begin{equation}
\omega_{\rm r}=\sqrt{\omega_{do}^{2}-\frac{\gamma^{2}}{2}}\approx
\omega_{do} \label{EqSec05e},
\end{equation}
and the sharpness of the resonance peak is determined by the quality
factor $Q$, defined as
\begin{equation}
Q=\frac{\omega_{do}}{\gamma}.
\label{EqSec05f}
\end{equation}
Larger $Q$ values correspond to sharper resonance peaks.

\begin{figure}
\centering
\includegraphics[width=12cm]{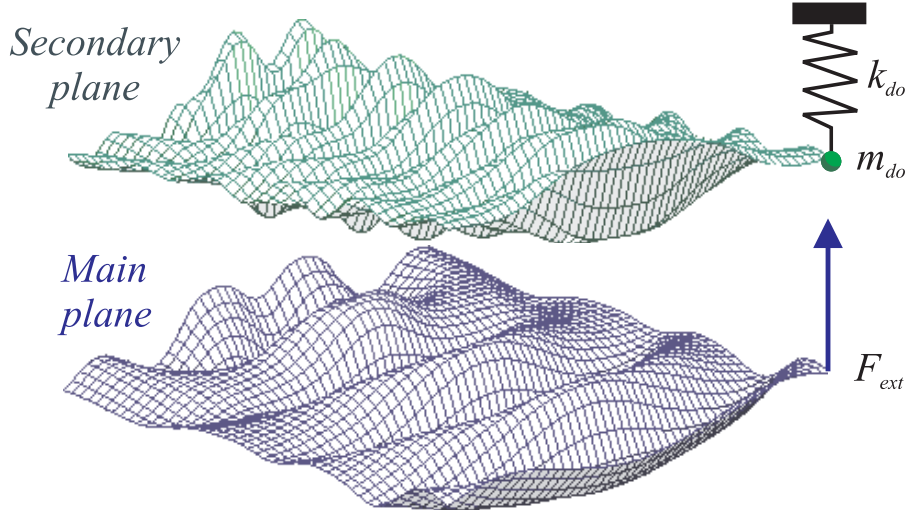}
\caption{\label{f15} Schematic diagram of the tuning filter
implementation.}
\end{figure}
The implementation of the \emph{TF} is similar to that of the
\emph{DF}. The simulation space where the multi-frequency
wavefield is evolving is considered as the main plane. Then, a
damped oscillator is associated to each point of the main plane,
forming a two-dimensional array of oscillators, called the
secondary plane. As schematized in Fig. \ref{f15}, the wavefield
amplitude at each point in the main plane is used to generate a
proportional force that is applied to the particle of mass
$m_{do}$ in the corresponding oscillator. The displacement of each
particle in the secondary plane is associated to the amplitude of
the filtered wavefield. The oscillators are tuned at the frequency
to be isolated, and then, the value of $\omega_{do}$ is selected
accordingly. Therefore, as many secondary planes as frequencies to
be extracted will be needed.

The main
advantage of the MFE is that the computing
time is significantly reduced compared with the SFE.
In principle, the MFE employs the same
time to complete the simulation of the wavefield as that used by
the SFE for just a single frequency. For a
spectrum containing $f_{\rm tot}$ discrete frequencies, the total
number of iteration cycles is reduced $f_{\rm tot}$ times with respect
to the number of iteration cycles required to process the same
signal when a sequential swept of $f_{\rm tot}$ single frequencies is
carried out. However, the advantage in speed is compensated by a
requirement of more memory to store the whole set of secondary
planes and their respective auxiliary variables, which also
introduces a slight delay in the computing time of each iteration
cycle of the MFE compared with that of the SFE. This delay
increases with the number of simultaneous frequencies explored in
the MFE. More precisely, the computing times for a single iteration
cycle within each excitation scheme are related by:
\begin{equation}
t_{\rm MFE}=t_{\rm SFE}\;(1+\delta_{\rm MFE}\, f_{\rm tot}),
\label{EqSec05f0}
\end{equation}
where $\delta_{\rm MFE}$ is the additional fraction of $t_{\rm
SFE}$ required by each frequency. In the case of the algorithm
implemented in this work, $\delta_{\rm MFE}\approx 0.03879$. In
general, $\delta_{\rm MFE}$ could vary according to the dynamic
memory allocation efficiency of the implemented algorithm.

Taking into account
(\ref{EqSec05f0}), to run a whole set of frequencies the MFE requires less
computing time than the SFE if
\begin{equation}
n_{cS}^{\rm tot} > (1+\delta_{\rm MFE}\, f_{\rm tot})\, n_{cM}^{\rm tot},
\label{EqSec05i}
\end{equation}
where $n_{c M}^{\rm tot}$ is the total number of iteration cycles
for the MFE and $n_{c S}^{\rm tot}=f_{\rm tot}\cdot n_{c}^{\rm tot}$ is the total number of iteration
cycles for the SFE.

While $n_{cS}^{\rm tot}$ is only determined by the size of the
simulation space and by $f_{\rm tot}$, in the case of the MFE
$n_{cM}^{\rm tot}$ is also conditioned by the required selectivity
of the \emph{TF}, which depends on $\Delta \omega$ and on $f_{\rm
tot}$. According to (\ref{EqSec05d}), the separation between
adjacent discrete frequencies should be $\Delta \omega \geq
\gamma$ to perform an adequate isolation of frequencies. Besides,
the \emph{TF} should work under the stationary regime, i.e., when
the non-tuned frequency oscillations have vanished. Since the time
employed by the oscillator to reach the stationary regime is
inversely proportional to $\gamma$, the time required by the
\emph{TF} to isolate the selected frequency basically depends on
the desired resolution $\Delta \omega$. Consequently, the value of
$\gamma$ should be adjusted according to the speed and precision
requirements of the simulation in each particular case, and this
determines the value of $n_{cM}^{\rm tot}$. According to the above
considerations, one could decide which scheme is more appropriate.

\subsection{Discretization requirements and validation}
\label{sec05b}

In the case of the optical spectrum ($\lambda \in$ [380 nm, 780
nm]) discretized in steps of 10 nm (41 frequency values),
the value of $n_{cM}^{\rm tot}$ required for an appropriate
separation of frequencies results to be large
since, in this case, $\Delta \omega = \gamma$ is small and then
the required quality factor is very high
(see eq. (\ref{EqSec05f})).
Moreover, to minimize the noise produced by the non-tuned components in the
tuned frequency wave, it is even convenient to choose $\gamma \ll \Delta \omega$,
and this implies a high value of $n_{cM}^{\rm tot}$.

The total number of cycles $n_{cM}^{\rm tot}$ required by the
\emph{TF} to extract a frequency could, in principle, be reduced
by increasing the time adapting constant $\tau_{n}$ (see eq.
(\ref{Eq00d})), i.e., by increasing $\omega_{d}$. As mentioned in
Section \ref{sec02}, the Nyquist-Shannon criterion imposes
$\omega_{d} < \pi$. However, it was found that for high values of
$\omega_{d}$ the \emph{TF} resonates at a frequency higher than
the tuned frequency. Then, to ensure that the \emph{TF} selects
the desired frequency, $\omega_{d}$ must satisfy $\omega_{d} \ll
\pi$, and therefore:
\begin{equation}
v_{d}\; \sigma_{p}=\frac{\omega_{d}\; c}{\omega} \ll \frac{\pi\;
c}{\omega}, \label{EqSec05j}
\end{equation}
for $\omega_{d}$ and $\omega$ being the maximum digitized and physical frequency
contained in the analyzed spectrum, respectively.

In order to quantify the performance of the \emph{TF}
independently from other characteristics of the simulation, we
define the relative error (in percentage) of the obtained
intensity as a function of the explored wavelength as:
\begin{equation}
\varepsilon_{r}(\lambda_{e})=100 \,
\frac{|I_{\rm MFE}(\lambda_{e})-I_{\rm SFE}(\lambda_{e})|}{I_{s}(\lambda_{e})|},
\label{EqSec05k}
\end{equation}
where $I_{\rm MFE}=(A^{\rm max}_{\rm M})^{2}$ and $I_{\rm SFE}=(A^{\rm max}_{\rm S})^{2}$ are
the intensities obtained with the MFE for
the wavelength $\lambda_{e}$ and the
SFE, respectively, and
$A^{\rm max}_{\rm M}$ and $A^{\rm max}_{\rm S}$
are the maximum amplitudes detected at the final stage of the
time evaluation (in which the \emph{TF} is in the stationary regime)
for the MFE and the SFE cases, respectively.

The exploration of the optical spectrum with a resolution of 10 nm
is an extreme situation for the MFE scheme because, as mentioned
above, for 41 simultaneous frequencies a very high quality factor
is needed, which leads to a very large number of iteration cycles
$n_{cM}^{\rm tot}$. This limitation can be overcame by exploring
the same amount of frequencies in several MFE stages having a
larger separation between frequencies, in such a way that
different frequencies are covered in each stage. In this manner,
the required quality factor of the \emph{TF} is greatly decreased
and therefore, $n_{cM}^{\rm tot}$ is also considerably reduced.
For example, the optical spectrum can be explored in five MPE
stages with a resolution of 50 nm ($f_{\rm tot}=9$). In this case,
an error of $\varepsilon_{r}=5.25$\% can be achieved with
$n_{cM}^{\rm tot}=50000$ cycles per stage, $\gamma=10^{-4}$
rad/cycle, $\sigma_{p}=19$ nm/pixel and $v_{d0}=0.5$ pixels/cycle.
Therefore, the total number of cycles required to explore the 41
frequencies will be $5\,n_{cM}^{\rm tot}=250000$. For $\delta_{\rm
MFE}\approx 0.03879$, the MFE scheme becomes more suitable than
the SFE in terms of computing time to explore the 41 optical
frequencies, if the simulation space becomes larger than $1454
\times 1454$ pixels. This example shows that for large simulation
spaces, the MFE (and the use of the \emph{TF}) represents an
advantage over the SFE, that can be implemented in any
conventional computer. The decision on which scheme (MFE or SFE)
is more suitable will in general depend on the maximum allowed
error in each particular case.

If we are just interested in the visualization of the field
associated to each frequency, less precision is required. In this
case, the whole set of frequencies can be analyzed in a single
stage, with a higher value of $\gamma$, which considerably reduces
the required number of iteration cycles.

\begin{figure}
\centering
\includegraphics[width=12cm]{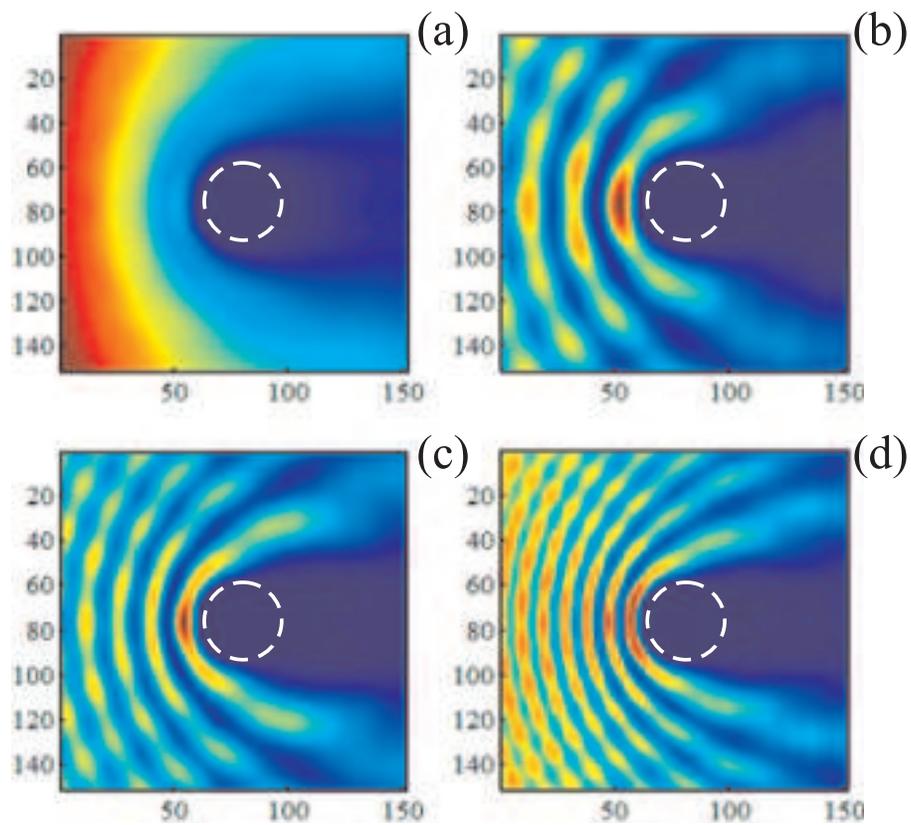}
\caption{\label{f16} Simulated intensity diagram of a
multi-frquency plane wave scattered by an opaque cylinder of a
diameter of 620 nm. (a) The multi-frequency wavefield; (b)
$\lambda$= 780 nm component; (c) $\lambda$= 570 nm component; (d)
$\lambda$= 380 nm component.}
\end{figure}

As an example, the field scattered by an opaque cylinder of
diameter 620 nm was simulated for an incident optical
multi-frequency plane wave ($\lambda \in$ [380 nm,780 nm], $\Delta
\lambda$ = 10 nm) with $\gamma=10^{-4}$ rad/cycle, $\sigma_{p}=20$
nm/pixel, $v_{d0}=0.5$ pixels/cycle, and a simulation space of
$150 \times 150$ pixels. Figure \ref{f16}(a) shows the resulting
multi-frequency intensity diagram for the scattered wavefield and
Figs. \ref{f16}(b)-(d), are the intensity diagrams for several
extracted components of wavelengths 780 nm, 570 nm and 380 nm,
respectively, for $n_{cM}^{\rm tot}=600$. The white dashed line
denotes the cylinder position. As can be observed, even for a very
low number of iteration cycles the \emph{TF} allows a clear
visualization of the wavefield of each frequency. This result can
be explained by taking into account that if $\gamma$ is relatively
high, the non-tuned frequency components are rapidly damped and
the filter would immediately be oscillating at its tuning
frequency, although the stationary regime has not been strictly
reached.

The possibility of getting a clear and rapid visualization of the selected frequency
components by means of the \emph{TF} could also be useful to
decouple the different frequencies that coexist in a multi-frequency
field of mechanical waves obtained experimentally.
A few thousands of frames of an experimental evolving
multi-frequency wavefield could be enough to decouple and
visualize the evolution of a single frequency component. This
feature also makes the \emph{TF} a valuable tool for visualization and
signal processing of experimental data.

\section{Near-to-far-field transformation method}
\label{sec06}

The method proposed in this work provides the near field
distribution of a wave interacting with a given object. However,
in many practical cases one is interested in the far-field response of an
illuminated structure. On the other hand, there are many
analytical methods that only provide the far field response of canonical
problems, that could serve for validation and for appropriately setting
the parameters of our simulation method. Therefore, in this Section we
explain how the far field is obtained from the near field.

The transformation method used to obtain the far field
from the near field is based on Green's Theorem
\cite{twenty-seven,thirty-eight}. The remarkable advantage of
this method is that it avoids the enlargement of the
simulation domain to obtain the
far-field information.

If the wavefield is represented by a complex phasor $Z(\textbf{r},t)=Z_R(\textbf{r},t)+
i\; Z_I(\textbf{r},t)$, according to Green's theorem \cite{twenty-seven} for a
fixed time we have:
\begin{equation}
Z_t(\textbf{r})=\oint_{C_{a}} [G(\textbf{r}|\textbf{r}')\;
\widehat{n}_{a}'\cdot
\nabla'Z_t(\textbf{r}')-Z_t(\textbf{r}')\;
\widehat{n}_{a}'\cdot \nabla'G(\textbf{r}|\textbf{r}')]\; dC',
\label{EqSec06a}
\end{equation}
where $Z_t(\textbf{r})$ is $Z(\textbf{r},t)$ evaluated at a fixed
$t$, $\textbf{r}'$ is the position of a source point over an
arbitrary contour $C_{a}$ enclosing the scatterer,
$\widehat{n}_{a}$ is the outward unit normal to the contour
$C_{a}$, $\textbf{r}$ is an observation point outside $C_{a}$, and
$G(\textbf{r}|\textbf{r}')$ is the Green's function, which in two
dimensions is given by the Hankel function
\begin{equation}
G(\textbf{r}|\textbf{r}')=\frac{i}{4}H_{0}^{(2)}(k|\textbf{r}-\textbf{r}'|),
\label{EqSec06b}
\end{equation}
with $i$ being the imaginary unit and $k = 2\pi/\lambda$ the
wavenumber.

To apply eq. (\ref{EqSec06a}) and calculate the far field, we need
to have the complex near field $Z_t(\textbf{r})$ for each pixel.
However, the present simulation method provides a real scalar
wavefield, and then, its imaginary part should be found. Taking
into account that
\begin{equation}
Z_\textbf{r}(t)=A_\textbf{r}(t) e^{-i \omega t}=A_\textbf{r}(t) \cos (\omega t)- i\;A_\textbf{r}(t) \sin
(\omega t),
\label{EqSec06c}
\end{equation}
where $A_\textbf{r}(t)$ is the amplitude of the phasor at a fixed
position $\textbf{r}$ and $\omega$ is the angular frequency in a
steady state (when $A_\textbf{r}(t)$ is constant), it is easily
verified that
\begin{equation}
Z_I(\textbf{r},t)=\frac{d}{dt}\{Z_R(\textbf{r},t)\}\;.   \label{EqSec06d}
\end{equation}

Since the simulation provides $Z_R(\textbf{r},t)$, we use
(\ref{EqSec06d}) to calculate $Z_I(\textbf{r},t)$ and therefore,
to build the phasor for each pixel that is introduced into eq.
(\ref{EqSec06a}) to calculate the far field.

\section{Application example}
\label{sec08}

\begin{figure}
\centering
\includegraphics[width=10cm]{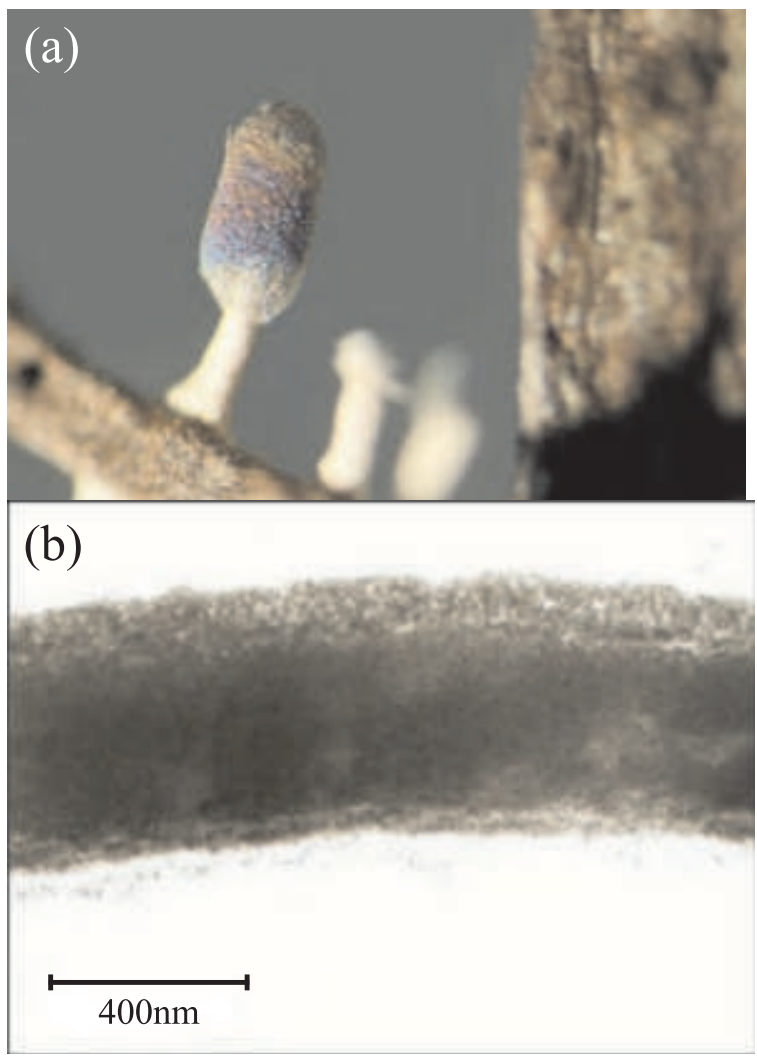}
\caption{\label{f17} (a) \emph{Diachea leucopoda} observed under
the optical microscope and (b) transmission electron microscope
image of the peridium cross section.}
\end{figure}

In this Section we show an application of the proposed simulation
and of the whole set of techniques developed in this work to
obtain the optical response of a highly complex structure of
biological origin. In particular, we evaluate the optical response
of the peridium --a transparent protective layer that encloses the
mass of spores-- of the \emph{Diachea leucopoda} (Physarales
order, \emph{Myxomycetes} class), which is a microorganism that
has a characteristic pointillistic iridescent appearance (see Fig.
\ref{f17}(a)). It has been recently demonstrated that this
appearance is of structural origin \cite{seventeen,thirty-nine}.
We use a transmission electron microscope (TEM) image of the
peridium cross section of the \emph{Diachea leucopoda}, shown in
Fig. \ref{f17}(b), to set the \emph{M} bitmap, which defines de
refraction index distribution by means of a linear conversion of
the grey levels of the negative image of Fig. \ref{f17}(b). The
grey level 0 (black) is associated to the lowest value of the
refraction index (equal to unity). The average refraction index of
the peridium was set to 1.78, which corresponds to a common value
found in biological tissues. The optical source is a gaussian beam
of width 2 $\mu$m. 41 optical frequencies were explored in the
range 380-780 nm with a spacing of 10 nm. The simulation space was
set to be of 280 $\times$ 230 pixels and $\sigma_{p}=15$ pixel/nm.
The width of the sample obtained from the TEM image is 1.68 $\mu$m
and its mean thickness is 550 nm. In order to avoid errors
produced by the finite size of the peridium image, we extended the
biological slab at both sides by planar homogeneous slabs whose
thickness is the average thickness of the actual image and whose
refraction index is its average refraction index.

\begin{figure}
\centering
\includegraphics[width=9cm]{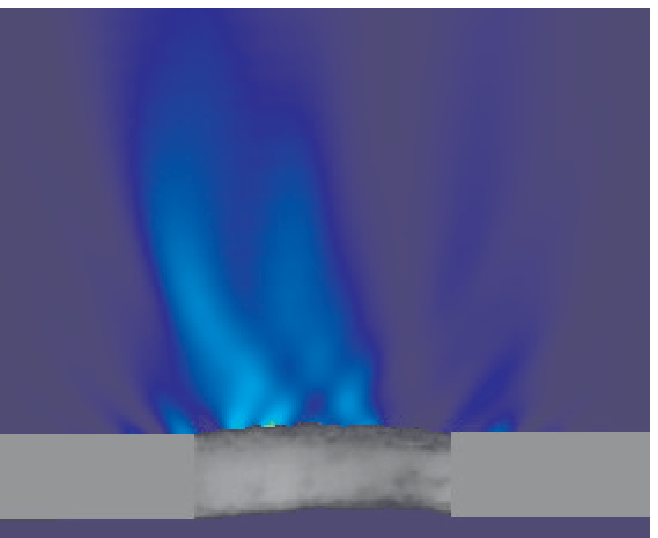}
\caption{\label{f18} Reflected near-field intensity diagram
produced by the peridium of the \emph{Diachea leucopoda} obtained
with the simulation method, for a wavelength of 380 nm.}
\end{figure}

\begin{figure}
\centering
\includegraphics[width=12cm]{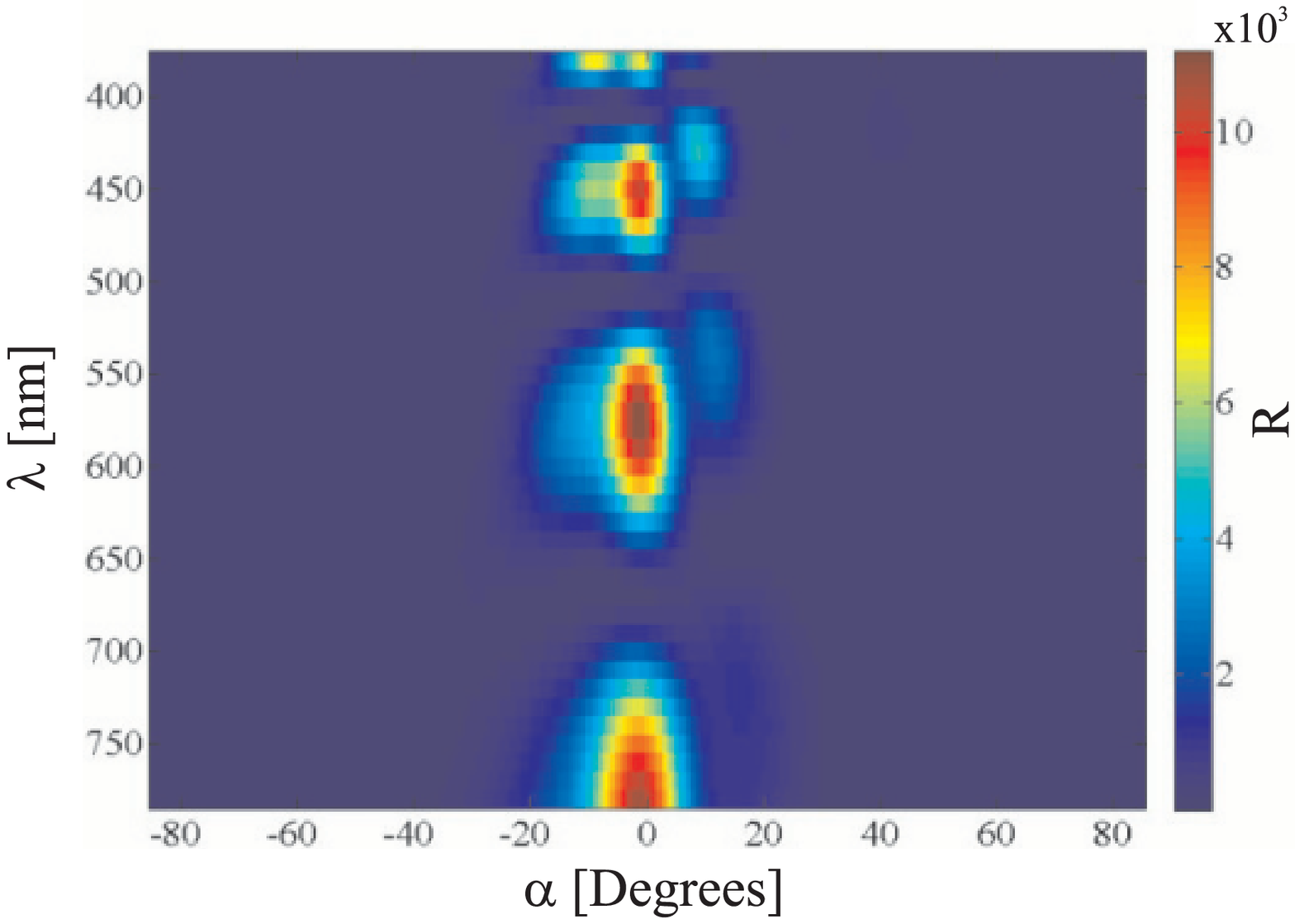}
\caption{\label{f19} Far-field reflectance (R) as a function of
the observation angle $\alpha$ and of the wavelength $\lambda$ for
the peridium of the \emph{Diachea leucopoda}.}
\end{figure}

Figure \ref{f18} shows the resulting near field distribution of
reflected intensity produced by the peridium cross section for a
wavelength of 380 nm, and Fig. \ref{f19} shows the total far field
reflectance as a function of the observation angle $\alpha$ and of
the wavelength $\lambda$ (the color scale bar is the same for
Figs. \ref{f18} and \ref{f19}).

\section{Conclusion}
\label{sec09}

In this paper we presented an electromagnetic wave simulation method
to obtain the time-varying fields for multi-frequency excitation
and arbitrary shapes and refraction index distribution of dielectric objects.
This method results particularly suitable for investigating the
electromagnetic response of photonic structures of biological origin,
that usually present a high degree of complexity in their
geometry as well as in the materials involved. To improve the performance
of the method from the point of view of computing time optimization and
space saving, we implemented a set of techniques that permit us controlling
and analyzing the propagating waves within the simulation. These methods include a
direction filter, that permits decoupling waves travelling in
different directions, a dynamic absorber, that prevents the
reflection of waves at the edges of the simulation space in order
to reproduce unbounded spaces, and a tuning filter that
allows multi-frequency excitation. We also adapted a near-to-far field method
to calculate the far field with a minimum use of computation time and allocation
space. Also, possible applications of these techniques have been suggested
to analyze time-varying experimental wavefields.
As an application example, we calculated the reflectance of the transparent cover
layer of a microorganism that exhibits iridescence, for multiple wavelengths and
observation angles.

In its present form, the proposed simulation method computes the
electromagnetic response in the case of transverse electric (TE,
electric field perpendicular to the plane of incidence) polarized
incident light. In order to fully simulate the electromagnetic
response of complex structures with traslational invariance, also
the transverse magnetic polarization mode (TM, magnetic field
perpendicular to the plane of incidence) should be included. This
would require an extension of the method to a fully vectorial
formulation, that is already under development. Another aspect in
which we are planning to work in the near future is the extension
of the simulation to deal with three-dimensional objects. As
stated above, most biological structures are highly complex and
require a 3D model to properly account for their electromagnetic
properties. The development of such a tool would constitute a
valuable contribution for the study of natural photonic structures.

\textbf{Acknowledgements}

The authors gratefully acknowledge partial support from Consejo
Nacional de Investigaciones Cient\'{\i}ficas y T\'ecnicas
(CONICET PIP 112-200801-01880) and Universidad de Buenos Aires
(UBA-20020100100533).

\end{document}